\documentclass[twocolumn]{aastex62}

\usepackage{apjfonts}

\newcommand\Ka{Ka LMC~1}

\shorttitle{Planetary Nebula in LMC Cluster NGC 1866}
\shortauthors{Bond et al.}

\newcommand{\ArIII}{\ion{Ar}{3}}

\newcommand{\HII}{\ion{H}{2}}
\newcommand{\NI}{\ion{N}{1}}
\newcommand{\NII}{\ion{N}{2}}
\newcommand{\OI}{\ion{O}{1}}
\newcommand{\OII}{\ion{O}{2}}
\newcommand{\OIII}{\ion{O}{3}}
\newcommand{\SII}{\ion{S}{2}}
\newcommand{\SIII}{\ion{S}{3}}

\newcommand{\Ha}{H$\alpha$}
\newcommand{\Hb}{H$\beta$}
\newcommand{\Teff}{T_{\rm eff}}

\newcommand{\kms}{{\>\rm km\>s^{-1}}}

\def\hei{\ion{He}{1}}
\def\heii{\ion{He}{2}}

\def\nii{\ion{N}{2}}

\def\oi{\ion{O}{1}}
\def\oii{\ion{O}{2}}

\def\sii{\ion{S}{2}}

\def\Gaia{{\it Gaia}}

\newcommand{\HST}{{\it HST}}

\def\mU{m_{\rm F336W}}

\def\mI{m_{\rm F814W}}

\begin{document}

\title{Serendipitous Discovery of a Faint Planetary Nebula in the Massive Young LMC Cluster NGC 1866
}

\author[0000-0003-1377-7145]{Howard E. Bond}
\affil{Department of Astronomy \& Astrophysics, Penn State University, University Park, PA 16802, USA}
\affil{Space Telescope Science Institute, 
3700 San Martin Dr.,
Baltimore, MD 21218, USA}

\author[0000-0001-5679-4215]{Nate Bastian}
\affil{Donostia International Physics Center (DIPC), Paseo Manuel de Lardizabal, 4, 20018, Donostia-San Sebasti\'an, Guipuzkoa, Spain}
\affil{IKERBASQUE, Basque Foundation for Science, 48013, Bilbao, Spain}

\author[0000-0003-3858-637X]{Andrea Bellini}
\affil{Space Telescope Science Institute, 
3700 San Martin Dr.,
Baltimore, MD 21218, USA}

\author[0000-0001-6604-0505]{Sebastian Kamann}
\affil{Astrophysics Research Institute,
Liverpool John Moores University,
IC2 Liverpool Science Park,
146 Brownlow Hill,
Liverpool L3 5RF, United Kingdom}

\author[0000-0001-9673-7397]{Mattia Libralato}
\affiliation{INAF - Osservatorio Astronomico di Padova, Vicolo dell'Osservatorio 5, Padova I-35122, Italy}

\author[0000-0002-4341-9819]{Florian Niederhofer}
\affil{Leibniz Institute for Astrophysics Potsdam (AIP),
An der Sternwarte 16,
14482 Potsdam, Germany}

\author[0000-0003-2451-739X]{Martin M. Roth}
\affil{Leibniz Institute for Astrophysics Potsdam (AIP),
An der Sternwarte 16,
14482 Potsdam, Germany}
\affil{Institut f\"ur Physik und Astronomie, Universi\"at Potsdam,
Karl-Liebknecht-Str.\ 24/25, 14476 Potsdam, Germany}
\affil{Deutsches Zentrum f\"ur Astrophysik, Postplatz 1, 02826 G\"orlitz, Germany}

\author[0000-0001-6213-4117]{Azlizan A. Soemitro}
\affil{Leibniz Institute for Astrophysics Potsdam (AIP),
An der Sternwarte 16,
14482 Potsdam, Germany}
\affil{Institut f\"ur Physik und Astronomie, Universi\"at Potsdam,
Karl-Liebknecht-Str.\ 24/25, 14476 Potsdam, Germany}

\correspondingauthor{Howard E. Bond}
\email{heb11@psu.edu}

\begin{abstract}

During an integral-field spectroscopic study of stars in the massive young open cluster NGC~1866 in the Large Magellanic Cloud, we serendipitously discovered a faint planetary nebula (PN). We designate it ``\Ka,'' and find that its location near the cluster center, along with the agreement of its radial velocity with that of the cluster, imply a high probability of membership in NGC~1866. The 200~Myr age of the cluster indicates that the PN's progenitor star had an initial mass of about $3.9\,M_\odot$. The integrated spectrum of \Ka\ shows strong emission lines of [\nii], consistent with it being a ``Type~I'' nitrogen-rich PN\null. The nebula exhibits a classical ring  morphology, with a diameter of $\sim$$6''$, corresponding to an advanced expansion age of about 18,000~yr. Archival images of NGC\,1866 obtained with the {\it Hubble Space Telescope\/} reveal a faint blue central star. Comparison of the star's luminosity with predictions from one set of theoretical post-asymptotic-giant-branch evolutionary tracks {(for single stars)} implies an age roughly consistent with the dynamical age of the PN, but the agreement with alternative modern tracks is much poorer. Analysis of the emission-line spectrum suggests considerable dust extinction within the nebula; however the central star possibly suffers little reddening because we may be viewing it nearly pole-on in a bipolar PN\null. Our accidental discovery was made using data that are not ideal for study of \Ka; we suggest several avenues of future targeted studies that would provide valuable and nearly unique new information for constraining models of late stellar evolution.

\null\vskip 0.2in

\end{abstract}

\section{Introduction: Planetary Nebulae in Star Clusters
\label{sec:intro} }

At the end of a low- or intermediate-mass star's nuclear-burning lifetime, it spectacularly transforms from a red giant to a compact hot star, surrounded by a glowing nebula---a planetary nebula (PN). This metamorphosis occurs when an asymptotic-giant-branch (AGB) star (initial mass $\sim$0.8--$8\,M_\odot$) sheds its outer layers, exposing a core that rapidly evolves to higher surface temperature. When the star reaches $\sim$30,000~K, UV radiation ionizes the surrounding ejecta, producing the PN---which lasts a few times $10^4$~yr before dissipating. The central star then cools down as a white dwarf (WD) for the rest of eternity. For recent reviews of PNe and their significance for astrophysics, see \citet{Kwitter2022}, \citet{Parker2022}, and \citet{Kwok2022}.

If a PN is a verified member of a star cluster, it becomes a unique laboratory for stellar evolution and nucleosynthesis. For such objects, we obtain the initial mass and composition of the PN's progenitor star---information that cannot be determined for PNe in the field. The absolute luminosity of the central star can be found using the cluster's known distance. This can be converted to a post-AGB age of the star, based on theoretical evolutionary tracks, which can be compared with the dynamical age of the PN derived from its radius and expansion velocity.
Additionally, the chemical composition of the PN, along with the progenitor mass, provides information about the dredge-up of processed material from the stellar interior---theoretically predicted to depend strongly on the initial mass of the progenitor (e.g., \citealt{Karakas2018, Henry2018, Kamath2023}). Moreover, the presence of a PN in a cluster establishes the cluster's main-sequence turnoff mass as a lower limit on the masses of stars that end their lives as WDs, rather than exploding as core-collapse supernovae. As \citet{Kwitter2022} emphasize, ``a PN located in a star cluster is a rare celestial gift.''

Unfortunately, Nature has been stingy in bestowing this cosmic largesse. For PNe that potentially belong to Galactic star clusters, the situation has been described in detail recently by \citet{BelliniPHR2025} and \citet{Fragkou2025}, and earlier by \citet{MoniBidin2014}, and we summarize here. 

Among Galactic globular clusters (GCs), three PNe \citep[see][]{Bond2020} have passed the membership criteria of lying within the tidal radius of the cluster, and having a radial velocity (RV), interstellar extinction, estimated distance, and central-star proper motion (PM) that are all in agreement with those of the host clusters. A fourth candidate, the PN JaFu~1, lies close on the sky to the Galactic GC Palomar~6, and was long considered a likely cluster member. However, \citet{Bond_JaFu12024} used multiple-epoch imaging of its central star with the {\it Hubble Space Telescope\/} (\HST\/) to show that its PM is discordant with that of the cluster.  

For Galactic open clusters (OCs), 
the number of PNe confirmed as members is likewise very small. There are over a dozen Galactic PNe that
lie tantalizingly near OCs on the sky, but almost all have failed one or more membership tests (see summaries in \citealt{Davis2019} and \citealt{Kwitter2022}). In the past few years, however, there have emerged two well-confirmed cases of PNe belonging to Milky Way OCs, and two more good candidates: (1)~IPHASX J055226.2+323724 (PN G177.5+03.1) has convincingly been shown to be a member of the OC M37, based on the RV of the PN and the PM of its central star given in \Gaia\/ Data Release~3  \citep[DR3;][]{2023A&A...674A...1G} both agreeing very well with those of the cluster \citep{Griggio2022, FragkouM37_2022}. (2)~The PN PHR~J1315$-$6555 was proposed as a member of the distant Galactic OC AL~1 by \citet{FragkouPHR2019}. Recently, \citet{BelliniPHR2025} confirmed its membership through a precise \HST\/ measurement of the PM of the PN's faint central star, which agreed with that of the cluster. (3)~The PN BMP~J1613$-$5406 is a candidate member of the OC NGC~6067 \citep{FragkouBMP2019, FragkouNGC6067_2022}. The PM of its central star based on \Gaia\/ data has a large uncertainty and is moderately discordant with cluster membership; however, an \HST\/ project (PI: H.E.B.) is underway that will reduce the uncertainties considerably and will allow a definitive PM membership test. (4)~The bright bipolar PN Hb~2 lies close to the Galactic OC NGC~2818.\footnote{The nomenclature of this PN and cluster in the literature has been inconsistent. The nebula was discovered visually in 1828 by James Dunlop, and in 1838 by John Herschel, who also noted the presence of the neighboring open star cluster. The PN or the cluster, or both, have been called NGC\,2818 by various authors. In order to distinguish them from each other, the cluster or the PN has sometimes been designated NGC\,2818A\null. Here we are using Hubble~2 (Hb~2) for the PN, since the nebula was first shown spectroscopically to be a PN by \citet{Hubble1921}, and we call the cluster NGC\,2818.}   Whether the PN actually belongs to the cluster has been debated for many years, but \citet{Fragkou2025} recently argued that the PN is a member, based on its RV and other considerations. Unfortunately, however, the \Gaia\/ PM for its faint nucleus has a large uncertainty, but is discrepant enough with that of the cluster to cast significant doubt on its cluster membership.

Outside the Milky Way, PNe known to belong to star clusters are similarly extremely rare. In the Local Group, a PN that is a likely member of an OC in M31 was identified by \citet{Bond2015} and analyzed spectroscopically by \citet{Davis2019}.  \citet{Larsen2006}, in a spectroscopic survey of OCs in star-forming spiral galaxies outside the Local Group, discovered three PN candidate members, lying in two OCs in M83 (4.5~Mpc) and one in NGC\,3621 (6.6~Mpc).

In this paper, we report the serendipitous discovery of a PN that is a candidate member of a massive young OC in the Large Magellanic Cloud (LMC)\null. We present details of the discovery, imaging and spectroscopy of the PN, identification and photometry of its central star, comments on its evolutionary status, and suggestions for future work that would provide further valuable information on this object.

\section{Discovery of a PN Superposed on the Massive LMC Cluster NGC\,1866} 

In a recent study, \citet[][hereafter K25]{Kamann2025} used observations with the Multi-Unit Spectroscopic Explorer (MUSE) integral-field spectrograph (IFS) on the European Southern Observatory (ESO) Very Large Telescope (VLT) in a study of stellar rotation in NGC\,1856 and NGC\,1866, two massive young star clusters in the LMC\null. K25 estimate ages of 300 and 200~Myr for these clusters, respectively. 

In addition to its utility for stellar spectroscopy, the large field of view, wide wavelength range, good spectral resolution, and high throughput all make MUSE an ideal instrument for discoveries, intentional or serendipitous, of faint emission-line sources.
MUSE observations have previously led to discoveries of nebulae in star clusters, both in the Galaxy \citep[e.g.,][]{Goettgens2019} and in the LMC \citep[for example,][]{Kamann2020}.
The superior sensitivity of MUSE for detection of low-surface-brightness emission-line objects such as supernova remnants, {\HII} regions, the diffuse interstellar medium (ISM), and PNe in nearby galaxies was reported in a pilot study on NGC\,300 by \citet{Roth+2018}, and exploited in the recent rejuvenation of the PN luminosity function (PNLF) as a tool for measurement of extragalactic distances \citep{Roth+2021, Jacoby+2024}. The high sensitivity to emission-line sources is due to the narrow bandwidth of monochromatic images provided by integral-field spectroscopy \citep{Roth+2004}.
For technical details of the MUSE instrument, see \citet{Bacon2014}.

\begin{figure*}
\centering
\includegraphics[width=0.8\textwidth]{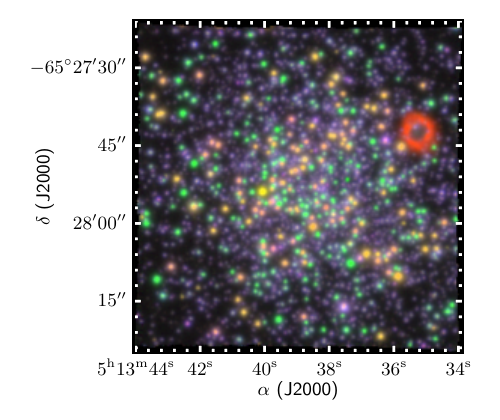} 
\hskip0.7in \hfill
\vskip-0.25in
\caption{
MUSE synthetic emission-line image of the LMC massive young cluster NGC~1866,  showing the newly discovered PN \Ka---the red elliptical ring on the northwest side of the cluster with a major axis of $\sim$$6''$ (1.3~pc at the distance of the LMC)\null. Color-coding is [\NII] $\lambda$6584 (red), \Ha\ (green),  and [\OIII] $\lambda$5007 (blue).  All wavelengths shifted to the radial velocity of the LMC\null. Image dimensions are $60''\times60''$. The numerous green objects are Be stars, bright at \Ha, in this young cluster that is rich in rapidly rotating upper-main-sequence stars. 
\label{fig:MUSEimage}
}
\end{figure*}

The MUSE data for NGC\,1866 discussed here, obtained on 2019 December~27, are available from the public ESO archive.\footnote{PI: S.~Kamann; \url{https://archive.eso.org/dataset/ADP.2020-01-09T23:43:57.105}} The  exposures were $4\times660$~s, all done within 1~hour of telescope time.
In the course of examining these data, S.K. was surprised to detect a previously unrecognized faint emission-line source, an apparent PN\null.  
Figure~\ref{fig:MUSEimage} shows a color rendition\footnote{An image based on the same data, but with a different color palette, was presented in the K25 discovery paper.} of the data cube, color-coded to represent emission lines of [\NII], \Ha, and [\OIII], as described in the figure caption. Here the PN candidate is seen on the northwest side of the cluster. A conspicuous slightly elliptical ring, dominated by [\NII] emission, is present, with a major axis of about $6''$ (corresponding to a linear size of 1.3~pc, if the PN is at the LMC distance). Following the standard naming convention\footnote{See the Hong-Kong/AAO/Strasbourg/H$\alpha$ (HASH) catalog of PNe at \url{http://hashpn.space/} \citep{Parker2016, Bojicic2017}.} for PNe, based on the discoverer's surname, we designate the nebula ``\Ka.''

\section{Nebular Imaging and Spectroscopy}


\subsection{Monochromatic Imagery}

Reductions of the MUSE data for NGC\,1866 were carried out with the standard ESO pipeline \citep{Weilbacher2020}, which performs the basic calibration steps (e.g., bias subtraction, spectrum tracing, wavelength calibration and correction to the solar-system barycenter) on a per-integral-field-unit (IFU) basis. The data from the 24 IFUs were then combined, corrected for spaxel-to-spaxel transmission variations using twilight flats, and flux-calibrated using a standard-star observation. These steps were repeated for every exposure obtained within a single observing block. Finally, data from the different exposures were combined and resampled to a regular 3-dimensional data cube. For further details on MUSE data reduction, we refer to K25.

Since there is significant overlap of the nebula by bright cluster stars, whose cores can be orders of magnitude brighter at emission-line wavelengths than the faint nebula, we used the residual data cube resulting from the PSF-fitting code {\tt PampelMuse} \citep{Kamann+2013}, as employed by K25, for further processing. The PSF fits to the stars are not  perfect, leaving residuals strong enough to affect monochromatic images, and thus biasing the spectrophotometry. In order to eliminate this systematic error, we localized
the residuals with small apertures and replaced them by the flux from adjacent apertures of identical radius, as a proxy for the true surface brightness at the locations of the stars.

\begin{figure*}
\centering
\includegraphics[width=1.0\textwidth]
{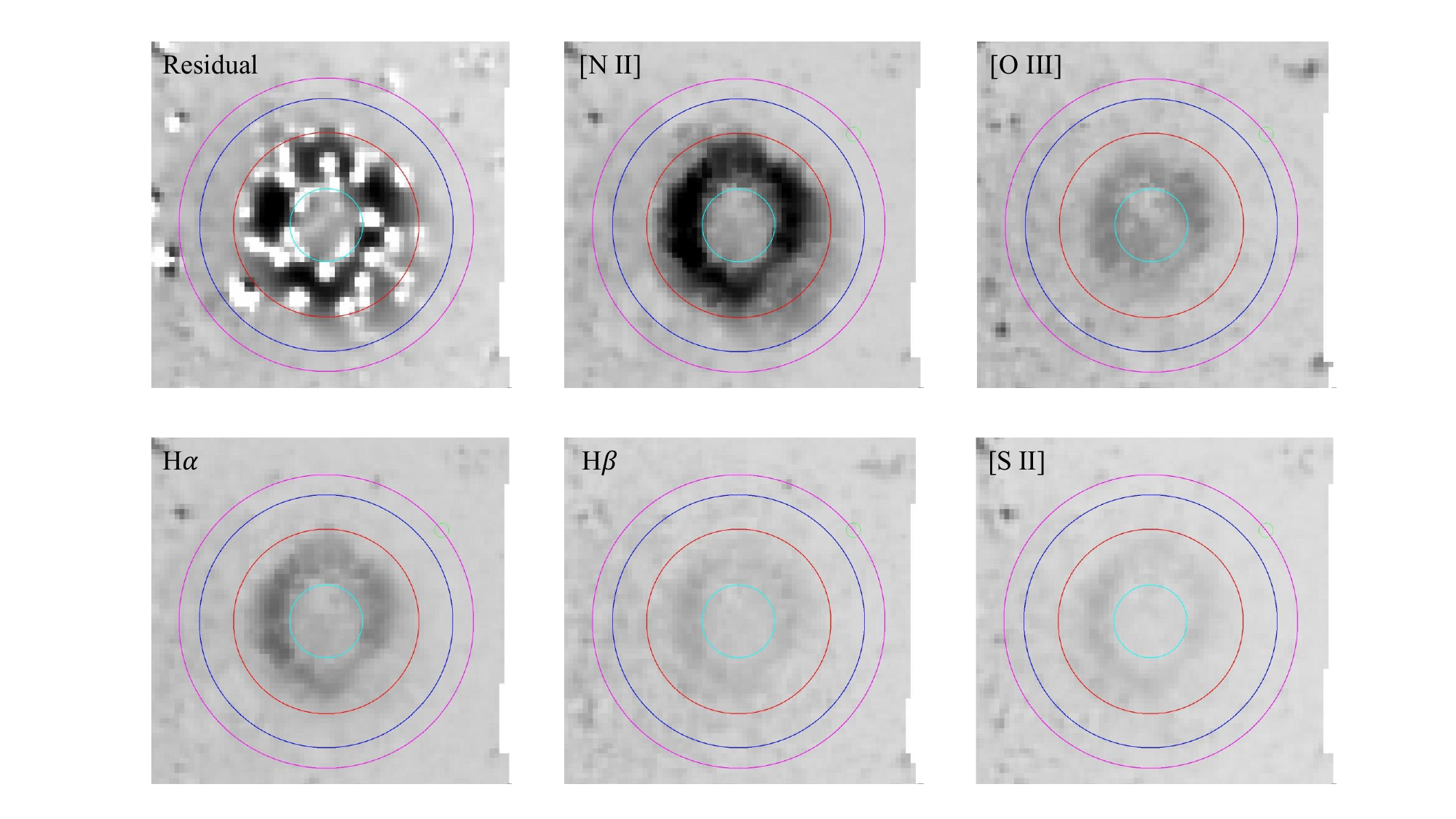} 
\hskip0.7in \hfill
\vskip-0.25in
\caption{Monochromatic images of \Ka\ after correction of stellar residuals at emission lines of [\NII]~$\lambda 6583$, [\OIII]~$\lambda 5007$, H$\alpha$, H$\beta$, and [\sii]~$\lambda 6716$, redshifted to the nebular radial velocity of +301.5~${\rm km\,s^{-1}}$. Each frame has dimensions of $14\farcs6\times14\farcs6$. For reference, the upper-left frame shows the [\NII] image before correction for superposed field stars. Here the white artifacts have amplitudes between $-1146$ and $-4854$ flux units per pixel. The greyscale stretch extends from $-70$ to 3000 units (asinh scaling). Morphological features are indicated with colored circles: cyan = inner bubble, red = rim, blue = shell. The sky annulus for {\tt DAOPHOT} background subtraction, used in creating the integrated spectrum of the nebula, is defined by the blue and magenta circles.}
\vskip 0.2in
\label{fig:monochromatic}
\end{figure*}

Figure~\ref{fig:monochromatic} shows five panels of emission-line maps that have been corrected for stellar residuals. For reference, an uncorrected map in [\NII]~$\lambda 6583$ is shown in the upper left panel.

The generic flux unit of calibrated MUSE data cubes is $10^{-20}$erg cm$^{-2}$ s$^{-1}$ per spaxel, whose size is $0\farcs2\times0\farcs2$. In faint parts of the PN, the surface brightness, e.g., in \Ha, is only of order 50 flux units per pixel above background, illustrating the challenge of photometric measurements.

The morphology of \Ka\ is best seen in the bright [\NII] emission-line image in the top-center panel in Figure~\ref{fig:monochromatic}. It is almost a textbook example of PN morphology, exhibiting an inner bubble, a bright rim, and a faint shell, according to the nomenclature of \citet{Schonberner2005}. The bubble, rim, shell, and a surrounding annulus for sky subtraction are indicated with cyan, red, blue, and magenta circles, respectively, in each panel of Figure~\ref{fig:monochromatic}.

\subsection{Integrated Nebular Spectrum \label{subsec:integrated_spectrum} }

\begin{figure*}
\centering
\includegraphics[width=0.9\hsize,bb=67 220 685 395,clip]
{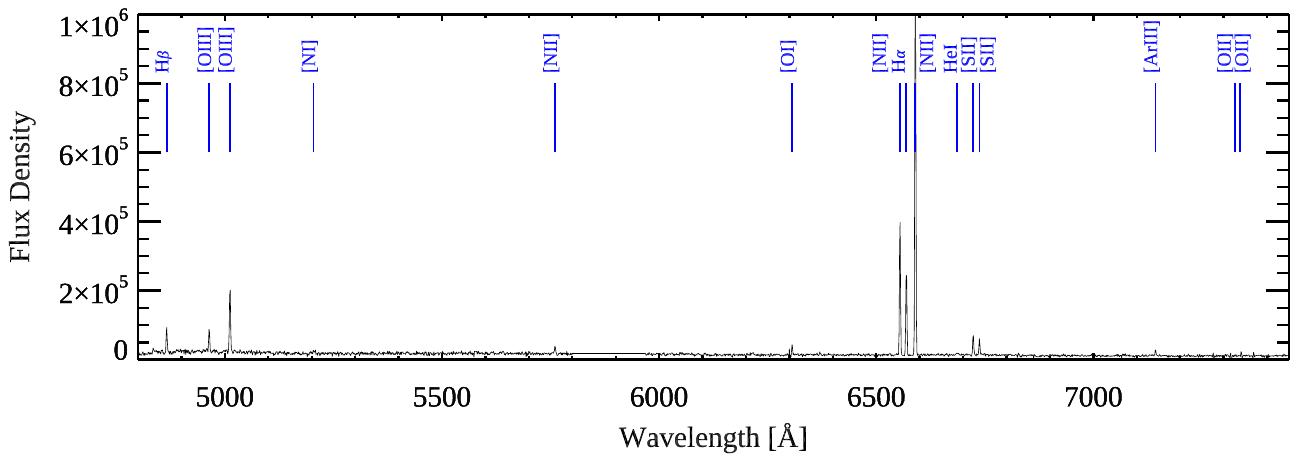} 
\includegraphics[width=0.9\hsize,bb=67 170 685 395,clip]
{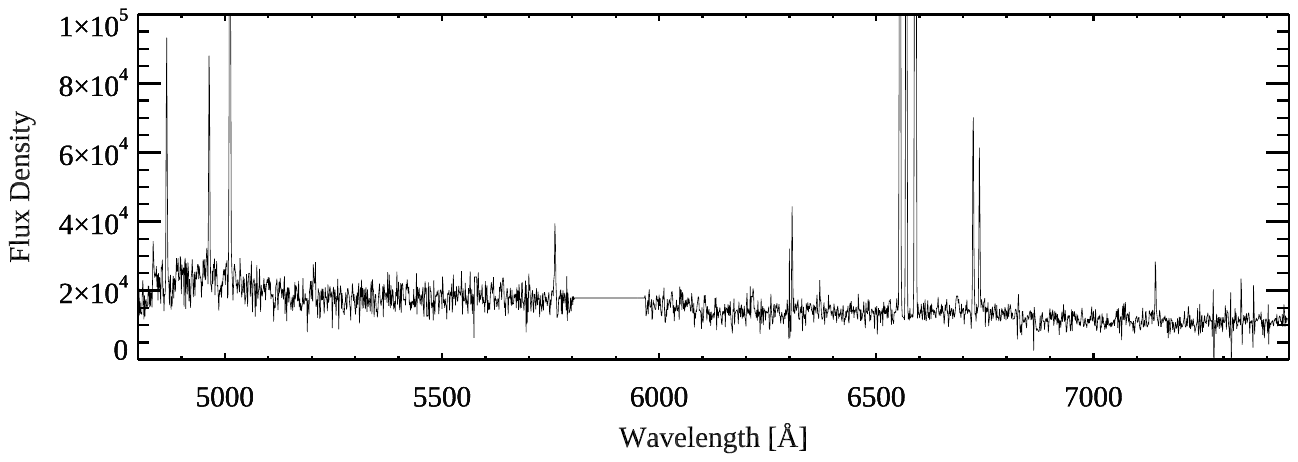} 
\hskip0.7in \hfill
\vskip-0.25in
\caption{Integrated spectrum of \Ka\ within an annular aperture that encloses its bright rim. Flux density is plotted in units of $10^{-20}$erg~cm$^{-2}$~s$^{-1}$~\AA$^{-1}$. The top panel shows the brightest emission lines, and in the bottom panel the scale is expanded to show the weak lines.
Selected emission lines are labeled in the top panel, and include H$\alpha$, H$\beta$, [\OIII]~$\lambda\lambda 4959,5007$, [\NII]~$\lambda 5754$, [\oi]~$\lambda 6300$, [\NII]~$\lambda\lambda 6548,6583$, [\sii]~$\lambda\lambda 6716,6731$, and [Ar\,III]~$\lambda 7136$. Wavelengths of a few lines of \hei, [\NI], and [\OII] that are weak or undetected are also marked.}
\vskip 0.2in
\label{fig:PNspectrum}
\end{figure*}

We extracted an integrated spectrum of \Ka\ from the MUSE data. To do this, we used {\tt DAOPHOT} \citep{Stetson1987} to perform aperture photometry at each layer of the data cube, with an aperture radius of $3\farcs8$ (19~pixels) in order to include the rim, and a sky annulus with inner and outer radii of $5\farcs0$ (25~pixels) and $6\farcs0$ (30~pixels), respectively. This resulted in a 1-dimensional spectrum in the interval  4700 to 9350~\AA, with spectral bins of {1.25~\AA} width, which is plotted in Figure~\ref{fig:PNspectrum}.

Using a Gaussian line-fitting tool written in {\tt IDL} \citep[see][]{Roth+2021}, we measured emission-line flux, line width, and the Doppler-shifted central wavelength for each of the detected emission lines in the PN spectrum. The results are listed in Table~\ref{tab:linefluxes}. 

\begin{deluxetable*}{lcccccc}
\tablecaption{PN Emission-line Fluxes
\label{tab:linefluxes}}
\tablewidth{0pt}
\tablehead{
\colhead{Ion} &
\colhead{$\lambda_{\rm rest}$} &
\colhead{Flux} &
\colhead{Flux Error} &
\colhead{$\lambda_{\rm centr}$} &
\colhead{FWHM} & 
\colhead{$V_{\rm rad}$} \\
\colhead{} &
\colhead{[\AA]} &
\colhead{}&
\colhead{} &
\colhead{[\AA]} &
\colhead{[\AA]} &
\colhead{[km s$^{-1}$]}
}
\startdata
H$\beta$     & 4861.32 &  21.3 & 1.3   &   4866.10  &	2.7  & $+294.7\pm3.3$  \\
{[\OIII]}    & 4958.92 &  19.2 & 1.1   &   4964.07  &	2.6  & $+311.2\pm3.3$  \\
{[\OIII]}    & 5006.85 &  59.4 & 1.1   &   5011.92  &	2.9  & $+303.4\pm3.0$  \\
{[\NII]}     & 5754.60 &  7.2  & 1.1   &   5760.33  &	3.0  & $+298.4\pm6.1$  \\
{[\OI]}      & 6300.32 &  7.6  & 0.46  &   6306.34  &	2.3  & $+286.3\pm6.1$  \\
{[\NII]}     & 6548.06 &  116  & 0.49  &   6554.60  &	2.7  & $+299.3\pm2.5$  \\
{H$\alpha$}  & 6562.78 &  78.0 & 0.50  &   6569.33  &	2.8  & $+299.1\pm2.5$  \\
{[\NII]}     & 6583.39 &  346  & 0.49  &   6590.04  &	2.7  & $+302.7\pm2.5$  \\
{[\SII]}     & 6716.42 &  17.6 & 0.86  &   6723.24  &	2.6  & $+304.3\pm3.2$  \\
{[\SII]}     & 6730.78 &  12.4 & 0.83  &   6737.63  &	2.5  & $+304.9\pm3.2$  \\
{[\ArIII]}   & 7135.80 &  4.35 & 0.64  &   7142.96  &	2.2  & $+300.6\pm6.5$  \\
{[\SIII]}    & 9068.90 &  5.82 & 0.47  &   9077.88  &	2.9  & $+300.6\pm6.6$  \\
\enddata
\tablecomments{Flux units are $10^{-16}$erg~cm$^{-2}$~s$^{-1}$. The listed line flux error is the formal uncertainty of the fit. $\lambda_{\rm rest}$: rest-frame wavelength. $\lambda_{\rm centr}$: Doppler-shifted central wavelength as determined from fit. FWHM: line width as measured from fit. $V_{\rm rad}$: barycentric radial velocity computed from measured central wavelength versus rest frame.}
\end{deluxetable*}

A striking feature of the integrated spectrum is the great strength of the [\NII] lines on either side of \Ha. This leads to a tentative classification of \Ka\ as a ``Type~I'' PN, as defined by \citet{Peimbert1978}. We return to an analysis and discussion of the nebular spectrum in Section~\ref{sec:nebular_analysis}.

\subsection{Absolute [\OIII] Flux}

The PN's [\OIII]~$\lambda$5007 emission-line flux of $5.94\times10^{-15}$erg cm$^{-2}$ s$^{-1}$ translates into an ``[\OIII] magnitude'' of $m_{5007} = 21.8$, as defined by \citet{Jacoby1989}. In the extremely deep planetary-nebula luminosity function (PNLF) of the LMC published by \citet{Reid+2010}, \Ka\ resides in the faint tail. It lies roughly 2.5~mag fainter than the completeness limit of the distribution---explaining why the PN has remained undiscovered until now---and almost 8~mag below the bright cutoff of the LMC's PNLF\null. This very low luminosity argues in favor of a highly  evolved PN whose central star is at a late evolutionary stage on its cooling track. The evolutionary status of the central star is discussed further below (Section~\ref{subsec:comparison_with_theory}).

\subsection{Radial Velocity and Cluster Membership}

The final column in Table~\ref{tab:linefluxes} lists the barycentric RV computed for each emission line in the spectrum of \Ka. The weighted mean RV from these measurements is $+301.5\pm5.8\,{\rm km\,s^{-1}}$. Barring an unlikely coincidence, this RV is consistent with the PN belonging to the LMC, rather than it being a member of the distant halo of the Milky Way superposed on the cluster.  


We now consider whether \Ka\ is a bound member of NGC\,1866, or belongs to the surrounding LMC field population. To estimate its membership probability, we used the RVs for stars in NGC\,1866 measured from the MUSE spectra that are presented in K25. To this data set we added RVs from \Gaia\/ DR3. Specifically, we searched the \textit{Gaia} archive for stars brighter than $G=17.5$ having RV measurements, and lying within a projected distance of 1~degree from the cluster center. We further required that the stars have measurements of parallax and photometry in all three \textit{Gaia} bands. This resulted in a sample of 1,742 stars. Combining the MUSE and \Gaia\/ data enables us to sample both the cluster kinematics and the LMC velocity field around the cluster.

We adopted the two-component kinematic analysis procedure presented in Section~3.2 of \citet{2023A&A...671A.106M}. Briefly, it is assumed that the observed kinematics can be modeled as a superposition of two populations. These are a cluster population, for which the radial stellar-density and velocity-dispersion profiles can be described by a \citet{1911MNRAS..71..460P} model; and a field population, for which a constant stellar surface density and velocity dispersion across the footprint covered by the data are assumed. Each star entering into the analysis is assigned a prior on its cluster membership probability, which is simply the relative projected number density of the Plummer profile at the projected distance of the star. To assign the priors, we adopted the cluster scale radius of $a=29\farcs24$ measured by \citet[][hereafter N24]{Niederhofer2024}.

We applied a Markov Chain Monte Carlo analysis to find the parameters that maximize the likelihood of the model, given the data. In our case, we optimized six parameters: the systemic velocity of the cluster, its central velocity dispersion, the scale radius of its dispersion profile, the mean velocity of the field population, the field velocity dispersion, and the fraction of field-to-cluster stars.\footnote{Note that  \citet{2023A&A...671A.106M} also modelled the rotation of their sample clusters. In our case, however, we assumed a non-rotating cluster.} The optimization was carried out using {\tt EMCEE} \citep{2013PASP..125..306F} and a chain consisting of 100 walkers, which were propagated for 500 steps, of which the first 250 were discarded as burn-in. In view of the low expected velocity dispersion of the cluster, we restricted our stellar sample to stars with RV measurement uncertainties below $\pm$$5\,{\rm km\,s^{-1}}$. We also removed all stars with RVs less than $+200\,{\rm km\,s^{-1}}$, which we considered to be foreground Milky Way stars. 

For the NGC~1866 cluster, we find a systemic RV of $v_{\rm sys}=+300.8\pm0.2\,{\rm km\,s^{-1}}$, a central dispersion of $\sigma_{\rm max}=5.7\pm1.0\,{\rm km\,s^{-1}}$, and a scale radius of $a=11^{+9}_{-4}$~arcsec. The field population has a mean velocity of $v_{\rm mean}=+291.8\pm1.6\,{\rm km\,s^{-1}}$ and a dispersion of $\sigma_{\rm field}=21.8\pm1.2\,{\rm km\,s^{-1}}$. Using these parameters, we were able to determine {\it a posteriori\/} membership probabilities for the sources in our RV sample.

\begin{figure}
\centering
\includegraphics[width=0.47\textwidth]{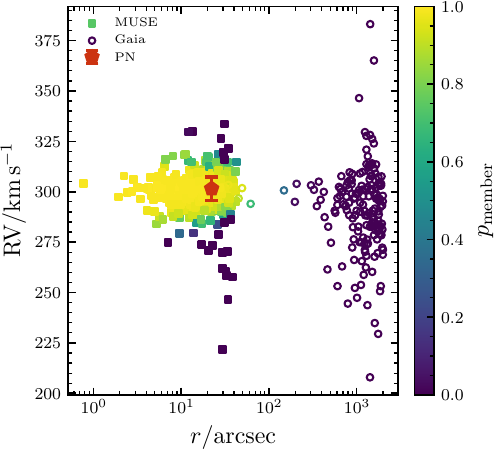}
\caption{
Barycentric radial velocities for stars in the vicinity of NGC\,1866, showing data from MUSE (filled squares), \Gaia\/ DR3 stars within 1~degree of the cluster (open circles), and the PN \Ka\ (red point with error bars), all plotted against angular distance from the cluster center. Stellar points are color-coded, as indicated by the color bar at the right, to represent their probabilities of cluster membership. See text for details of the data and probability calculations.
\label{fig:los_velocities}
}
\end{figure}

The results are shown in Figure~\ref{fig:los_velocities}, where the RVs are plotted against radial distances of the objects from the center of the cluster. The cluster membership probabilities for each star are color-coded as indicated by the color bar at the right side. Filled squares represent RVs measured in the MUSE sample, and open circles are RVs from the \Gaia\/ DR3 1~degree sample. The figure illustrates that for the vast majority of sources, their affiliation with either NGC~1866 or the LMC field is constrained to high confidence. For \Ka\ itself, plotted as a red point with error bars, we find a high probability of 98.1\% that it belongs to NGC~1866.

\subsection{Dynamical Age \label{subsec:dynamical_age} }

Assuming that the PN is at the distance of the LMC, we can estimate its dynamical age, based on its observed linear radius and an adopted expansion velocity. For a typical value in PNe of $v_{\rm exp} \simeq 35$~km\,s$^{-1}$ \citep[e.g.,][their Section 7.3]{Kwitter2022}, we find the following estimated age of the PN:

$$ \tau_{\rm dyn} \simeq 18000 \left({{r_{\rm neb}}\over{0.65\,{\rm pc}}}\right) \left({{35\,\kms}\over{v_{\rm exp}}}\right) \,\rm yr \, .$$

An age as large as $\sim$18,000~yr is, again, consistent with the PN being in an advanced stage of evolution.
In should be borne in mind, however, that the assumption of a constant expansion velocity is a simplication. As shown with hydrodynamical models for the expansion of PNe, one measures different expansion velocities for different ions, e.g., [\OIII] versus [\NII], and expansion velocities 
also vary over time \citep[e.g.,][]{Schonberner2005b}. As a rule of thumb, the dynamical age is shorter than an estimate assuming a typical constant $v_{\rm exp}$.


\null\medbreak

\section{\HST\/ Imagery Reveals the Central Star \label{sec:HSTimagery} }

NGC\,1866 is a massive [$\log(M/M_\odot)\simeq4.91$; \citealt{McLauoghlin2005}] young cluster located in the northern spiral structure of the LMC disk.
The Wide Field Camera~3 (WFC3) on \HST\/ has been used in three programs to obtain imagery of the cluster. Two were carried out in 2016 (GO-14069, PI: N.~Bastian; and GO-14204, PI: A.~Milone). Filters used were broad-band F336W,  F438W, F555W, and F814W, along with narrow-band F343N\null. \HST/WFC3 imaged the cluster again (F814W only) in 2022, under program GO-16748 (PI: F.~Niederhofer).



\begin{figure*}[!ht]
\centering 
\includegraphics[width=0.9\textwidth]{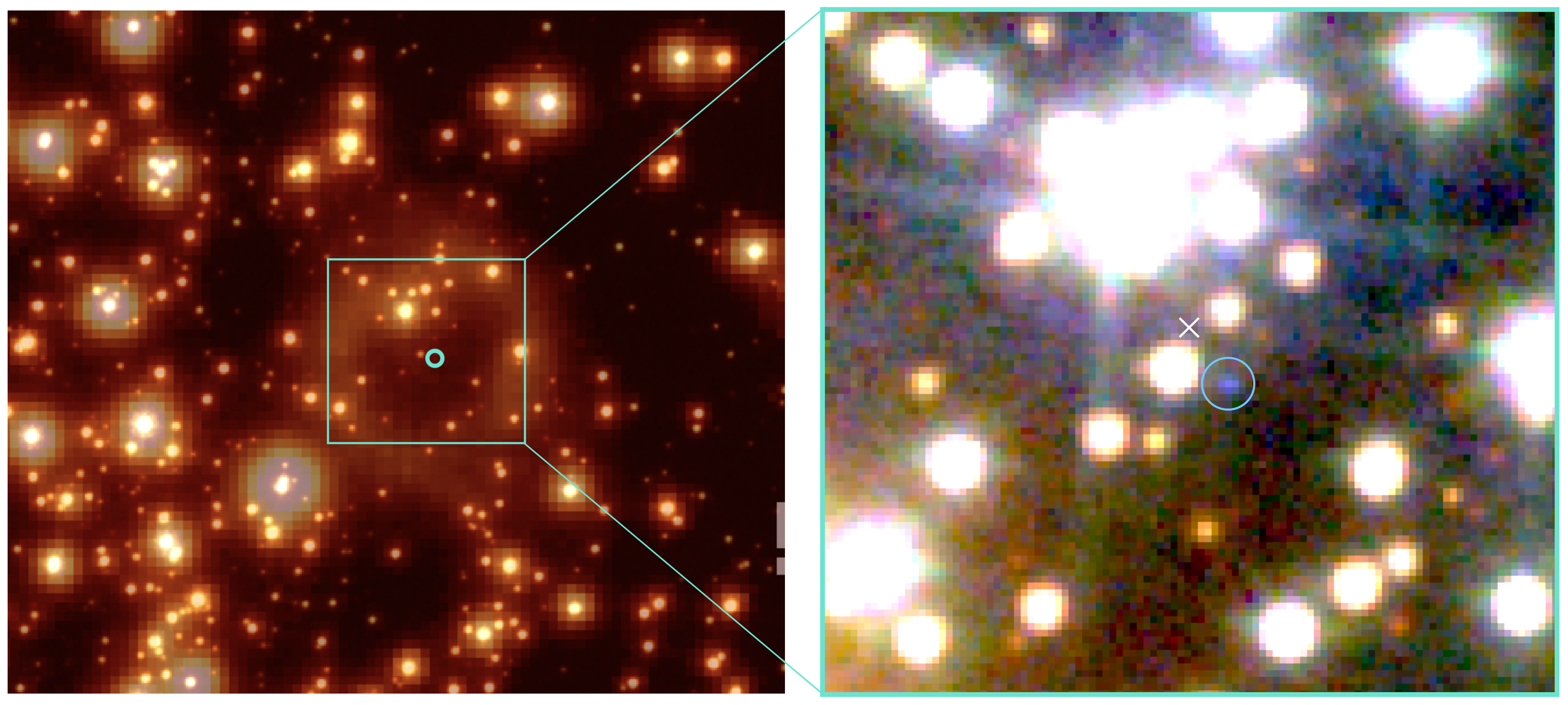}
\caption{
{\it Left:} Image of the site of the PN in NGC\,1866 from a stack of \HST/WFC3 F336W images (stars with sharp PSFs), superposed with a partially transparent MUSE data cube in \Ha. A small cyan circle marks a faint blue star near the center of the PN\null. 
{\it Right:} \HST/WFC3 color image in F336W, F438W, and F814W, zooming in on the cyan square in the left panel. Height of this frame is $4\farcs4$. A white X marks the geometric center of the nebula. Nearby, inside the circle, is a faint, very blue star, the likely nucleus of the PN\null. See text for details of these frames.
\label{fig:muse+hst}
}
\vskip0.2in
\end{figure*}

In Figure~\ref{fig:muse+hst} the left panel shows an enlargement of a stack of \HST/WFC3 images in the F336W filter centered on \Ka, producing the sharp stellar images. We registered a slice of the MUSE data cube at the wavelength of \Ha\ at the velocity of the LMC with the WFC3 frame, and superposed it with sufficient transparency to retain the WFC3 stellar images. \Ka\ is visible in the MUSE data as a faint elliptical ring. A small cyan circle marks the location of a faint blue star seen in the \HST\/ images. 

The right panel in Figure~\ref{fig:muse+hst} zooms in on the area marked by the cyan square in the left panel. It shows a color rendition of the \HST\/ frames in F336W (blue), F438W (green), and F814W (red). The white X marks the geometric center of \Ka. Nearby, inside a circle, is a faint ($V\simeq26.8$) and very blue star, which we identify as the PN nucleus (PNN)\null. 

Table~\ref{table:central_star} gives celestial coordinates and magnitudes for the  PNN\null. 
The astrometric frame is based on sources in common with the \Gaia\/ DR3 catalog (see Section~2.3 of N24 for details).
The stellar photometry given in the table was measured as described in the next section.

\begin{deluxetable}{lc}[h]
\tablecaption{Parameters for the Candidate Central Star of \Ka
\label{table:central_star} }
\tablewidth{\textwidth}
\tablehead{
\colhead{Parameter}
& \colhead{Value}
}
\startdata
RA (J2000) & 5:13:35.561\\
Dec (J2000) & $-65$:27:41.23 \\
$\mU$ [mag] & $24.76 \pm 0.10$ \\
$\mI$ [mag] & $26.71 \pm 0.60$ \\
\enddata
\tablecomments{Coordinates are in the \Gaia\/ DR3 frame for equinox J2000, epoch 2016.0. Magnitudes are on the Vega scale. Quoted magnitude errors are internal only and are approximate. See text for details.}
\end{deluxetable}


\section{Stellar Photometry}

\subsection{Cluster Members \label{subsec:cluster_members} }

Stellar photometry for members of NGC~1866, derived from \HST/WFC3 data for the cluster, has been analyzed and interpreted by several teams. These include \citet{Milone2017}, \citet{Milone2018}, \citet{Costa2019}, \citet{Milone2023}, N24, \citet{Ettorre2025}, and K25. These studies have generally found the cluster to have an age of about 200~Myr, a subsolar metallicity of $\rm[Fe/H]\simeq-0.4$, and modest interstellar reddening of $E(B-V)\simeq0.12$. 

\begin{figure}[!]
\centering 
\includegraphics[width=0.47\textwidth]{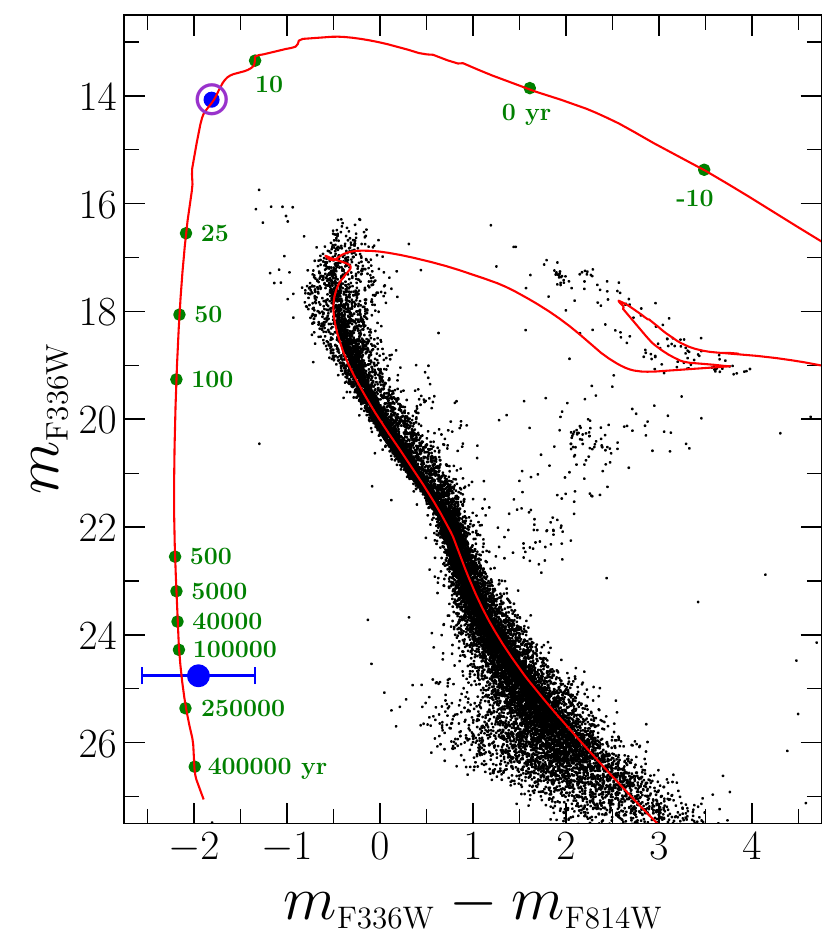}
\caption{
\HST/WFC3 color-magnitude diagram for NGC\,1866 (black dots). Superposed is a MIST isochrone (red line) for a distance modulus of $(m-M)_0=18.30$, age of 200~Myr, metallicity of $\rm[Fe/H]=-0.4$, reddening $E(B-V)=0.12$, and a single rotation velocity of 0.4 times critical. The blue filled circle with error bars marks the observed magnitude and color of the PN's central star. Timestamps along a MIST post-AGB evolutionary track for the PNN are marked, starting from the time at which its temperature reached $\log(\Teff/{\rm K})=3.85$. The circled point at the upper left marks the time at which the star reached $\Teff=30,000$~K, nominal onset of photoionization of the PN\null. See text for details of this figure.
\label{fig:cmd}
}
\end{figure}

Figure~\ref{fig:cmd} plots a color-magnitude diagram (CMD) for the cluster: apparent $m_{\rm F336W}$ magnitude versus $m_{\rm F336W}-m_{\rm F814}$ color. Here the black dots represent stars from the photometric and astrometric catalog for NGC\,1866 published by N24. As discussed by N24, the photometry for individual stars has been adjusted for a small amount of differential extinction to a constant amount, using techniques described by \citet{Milone2023}. 
We plot only well-measured stars, using the same selection criteria as in N24; these are based on photometric and astrometric quality cuts provided by the software package described in the next subsection.

\subsection{Central Star \label{subsec:central_star} }

Unfortunately, our candidate PNN is not contained in the N24 catalog. N24's reductions were focused on measuring high-precision proper motions of stars in NGC\,1866. Since the second-epoch data for this purpose were obtained only in F814W, the star-finding step of the process was set up to maximize the number of measured stars in the red band. This resulted in the omission of the blue PNN, which is so faint at F814W that it failed the detection algorithm. Therefore, to obtain photometry for it from the archival WFC3 frames, we proceeded as follows.

The N24 reductions employed the \texttt{KS2} software package (see \citealt{Bellini2017} for a detailed description), which provides three different approaches for measuring stellar properties. The first one, method~1, produces the best astrometric measurements, and was used to create the data catalog published by N24. The other two approaches are more suitable for photometry of faint stars, at the expense of poorer position estimates. In particular, \texttt{KS2}'s method~3 is the one that provides the best photometry for the faintest sources.\footnote{Method~1 measures source positions and fluxes via PSF fitting. Method~3 is a method of forced photometry: it adopts the stellar position determined in the finding stage. It then uses that position and the most significant four pixels, weighted by the expected values of the PSF at those pixels, in order to determine the best flux estimate for the source.}

We re-ran the \texttt{KS2} reduction package on the same first-epoch WFC3 data used by N24, but performed the star-finding step on the F336W images only. This approach allowed the package to find the central star in these frames. Its flux was then measured using method~3. Time-dependent zero-points were adopted as described by \citet{Calamida2022}, yielding a Vega-scale F336W magnitude. To this we applied a correction for differential extinction, determined as described in the previous subsection, of $-0.056\pm0.008$~mag. The result is listed in the third line in Table~\ref{table:central_star}; the quoted error is internal only.  

However, the PNN is too faint in the redder frames to be found directly by the reduction software. Instead, we had to rely on forced photometry, carried out at the F336W-derived position. Even then, the signal in the available F438W and F555W images was too weak to provide a useful measurement. We did successfully measure the flux with method~3 in three first-epoch F814W frames with long (678~s) exposure times, from the GO-14204 program described in Section~\ref{sec:HSTimagery}. The resulting F814W magnitude is listed in the fourth row of Table~\ref{table:central_star}. A differential-reddening correction of $-0.021\pm0.003$~mag has been applied.

Because of the failure of the star-finding algorithm, the forced photometry in F814W will have a large uncertainty. In order to provide a more reliable error estimate, we inserted 10,000 artificial stars into the individual frames
with magnitudes within $\pm$0.25~mag of the measured value. Here we followed the prescription given in Section~6 of \citet{Bellini2017}. Each artificial source was added one at a time, measured together with its neighbors, and then removed from the images, so that the inserted sources never interfered with each other.

An artificial source is considered successfully recovered if its measured position is within $\pm$0.5~pixel of its input position, and if its measured flux is within a factor of two from its input flux. Since the finding stage for the PNN was performed on F336W exposures only, we applied the same rules to artificial stars. We then further limited the analysis of photometric errors to artificial sources whose input magnitudes are within 0.1 mag of the PNN magnitude. For them, the photometric RMS of the residuals around the mean was 1.05~mag in F814W\null. Dividing this by the square root of three (the number of individual measurements in the F814W frames) yields the uncertainty given in the fourth row of Table~\ref{table:central_star}. However, it should be noted that photometric errors coming from artificial-star tests are always underestimated since, e.g., artificial sources are added and then measured with perfect knowledge of the PSF, while in reality we cannot know the true shape of the PSF with infinite precision. 



The resulting magnitude and color of the PNN are plotted as a filled blue circle with error bars in Figure~\ref{fig:cmd}. Note that the formal error in the F336W magnitude is smaller than the radius of the plotting symbol. 

We caution that our results may not have been fully corrected for the effects of a degraded charge-transfer efficiency (CTE) in the WFC3 detectors. 
Although our analysis did include the latest pixel-based CTE correction algorithm \citep{Anderson2021}, it may still suffer from residual CTE flux losses. Recently, \citet{Kuhn2024} computed a set of corrections for such losses, based on observation date, source flux, distance from the readout amplifiers, and total background. If we applied this correction, the PNN would become $\sim$0.1--0.3~mag brighter in F336W and $\sim$0.05--0.2 mag bluer in color than shown in our CMD.

However, these values should not be  applied blindly,  for multiple reasons. First, the \citet{Kuhn2024} CTE corrections represent maximum losses, meaning that the corrections to our results could be smaller. Second, our data reduction involves first- and second-pass photometry (see N24 for details), and we do not yet know how these corrections propagate through the entire reduction.  Lastly, our photometry is registered to that of one image using bright stars, each of which suffers from a different CTE loss, according to its position on the detector. Thus, a comprehensive correction would require a more in-depth analysis of the effect, which is beyond the scope of this paper. 

\null\medbreak 

\section{Stellar Evolution}

\subsection{Evolution in the Cluster \label{subsec:cluster_evolution} }

The CMD of NGC 1866 in Figure~\ref{fig:cmd} shows an extended main-sequence turn-off and a split in the upper main sequence. These phenomena are typical features in young Magellanic Cloud star clusters, and are caused by the effects of a range of stellar rotation from near zero to close to breakup, as well as a range of viewing angles of the stars (e.g., \citealt{Bastian2009}, \citealt{Wang2023}, \citealt{Kamann2023}, and the references cited in Section~\ref{subsec:cluster_members}).

Superposed on the CMD in Figure~\ref{fig:cmd} is a red line showing an isochrone, which we obtained using the MESA Isochrones and Stellar Tracks (MIST) Web Interpolator.\footnote{\citet{Paxton2011}, \citet{Dotter2016}, and \citet{Choi2016};  \url{https://waps.cfa.harvard.edu/MIST/interp_isos.html}} The parameters for this isochrone are the age, metallicity, and reddening given in the first paragraph of Section~\ref{subsec:cluster_members}, with a stellar rotation of $v/v_{\rm crit}=0.4$. A distance modulus of $(m-M)_0=18.30$ has been assumed (see N24 and references therein). This isochrone gives a reasonable fit to the main sequence of NGC\,1866, except for the extended and split main-sequence turnoff region, since only a single rotation and viewing angle was assumed for the theoretical isochrone. 

\subsection{Post-AGB Evolution of the Central Star \label{subsec:postAGB_evolution} }

The isochrone sequence running along the top of the CMD in Figure~\ref{fig:cmd}, and then descending almost vertically, represents stars in the final post-AGB stage. These objects evolve first to higher temperatures at nearly constant bolometric luminosity, and then begin their descent of the WD cooling track. The hot post-AGB stars lying on the MIST isochrone at $\mU\simeq14$ had an initial mass of $\sim$$3.88\,M_\odot$, and those at $\mU\simeq26.5$ an initial mass of $\sim$$3.91\,M_\odot$. The final mass for these stars, upon reaching the beginning of the WD phase, is about $0.86\,M_\odot$.

Stellar evolution in the post-AGB phase is extremely rapid, especially during the initial transition from the AGB to high temperatures. Thus, in this stage the isochrone shown in Figure~\ref{fig:cmd} is nearly identical to an evolutionary track for a star with an initial mass of $\sim$$3.9\,M_\odot$. To illustrate the post-AGB evolutionary timescale, we used the MIST website to obtain a theoretical track for a star with an initial mass of $3.90\,M_\odot$ and a metallicity of $\rm[Fe/H]=-0.4$; the MIST interpolator converts the theoretical values to WFC3 magnitudes for a reddening of $E(B-V)=0.12$, and we again assume a distance modulus of $(m-M)_0=18.30$. Following, e.g., \citet[][hereafter MB16]{MillerBertolami2016}, we define the onset of the post-AGB stage as the time at which the star's temperature has risen to $\log\,(\Teff/{\rm K})=3.85$. This point on the track is marked ``0 yr'' in Figure~\ref{fig:cmd}. Successive positions along the track of the remnant are labeled with their post-AGB ages, up to a final point at $\sim$400,000~yr.\footnote{The MIST $3.9\,M_\odot$ track actually ends at a post-AGB age of $\sim$85,000~yr; the last few age points plotted in Figure~\ref{fig:cmd} are based on an approximate linear extrapolation.} Shortly after the 10~yr point, the star reaches an effective temperature of 30,000~K, marking the nominal onset of photoionization of the PN; an open circle surrounding a small filled circle marks this point in the figure.

The timestamps on the post-AGB track demonstrate the extraordinarily fast initial evolution of the star after it leaves the AGB\null. Its traversal across the top of the CMD, transforming from a luminous red giant to a hot PNN, takes just a few dozen years. From there, the star fades by nearly 5.5~mag in F336W in the first 100~yr, and another 3~mag at a post-AGB age of $\sim$500~yr. At this point, however, the evolution slows down dramatically, as the star approaches its final WD radius. 


\null\medbreak

\subsection{Confrontation with Theoretical Evolutionary Tracks \label{subsec:comparison_with_theory} }

Knowing the position of the central star in the CMD, and having an approximate dynamical age for the PN of $\sim$18000~yr (see Section~\ref{subsec:dynamical_age}), we are in a position to use \Ka\ as a testbed for theoretical models of post-AGB stellar evolution. 

We consider two modern sets of models that trace stellar evolution all the way from the main sequence to the post-AGB phase. These are the widely cited evolutionary tracks of MB16, and tracks generated by the MIST tool referred to in Section~\ref{subsec:cluster_evolution}. The MB16 tracks have been calculated only for a relatively small set of initial stellar masses and compositions. The MIST website, through interpolation algorithms, allows any choice of initial mass and metallicity. 

In order to compare the tracks with observations, the theoretical quantities have to be translated into absolute stellar magnitudes in a given photometric system---in our case magnitudes in the WFC3 F336W and F814W filters. Unfortunately, for the MB16 tracks, a translation to WFC3 magnitudes has to our knowledge not been carried out (confirmed by M.~Miller Bertolami, priv.\ comm.). 

We therefore adopted a simple procedure to perform this step. We assume a blackbody spectral-energy distribution (SED) at a given central-star effective temperature, $\Teff$. This approximation is justifiable at optical wavelengths for PNNi with temperatures in the range of  $\Teff\simeq100,000$~K;  see, e.g., the PNN SEDs plotted in \citet{Reindl+2024}. We normalize to the luminosity of the Sun, and multiply by the model luminosities. For the effective wavelength of the F336W filter, 3355~\AA, the flux of the model can be read from the Planck curve, and scaled to the distance modulus of $(m-M)_0=18.30$. Finally, the Vega magnitude at the same wavelength is obtained from the calibration spectrum provided by \citet{Bohlin2020}, allowing normalization to the Vega-based scale. The F814W magnitude is obtained in analogous ways. These magnitudes are intrinsic, so observed magnitudes must be dereddened in order to make comparisons. 

The MIST tool performs translations to the WFC3 system, using SEDs from several libraries of model stellar atmospheres. These results were used for the reddened MIST isochrone plotted in Figure~\ref{fig:cmd}. However, for the sake of consistency in a direct comparison of the two sets of tracks, we performed the same blackbody conversions for the MIST tracks that we applied to the MB16 tracks.

\begin{figure}[h]
\centering 
\includegraphics[width=0.47\textwidth]{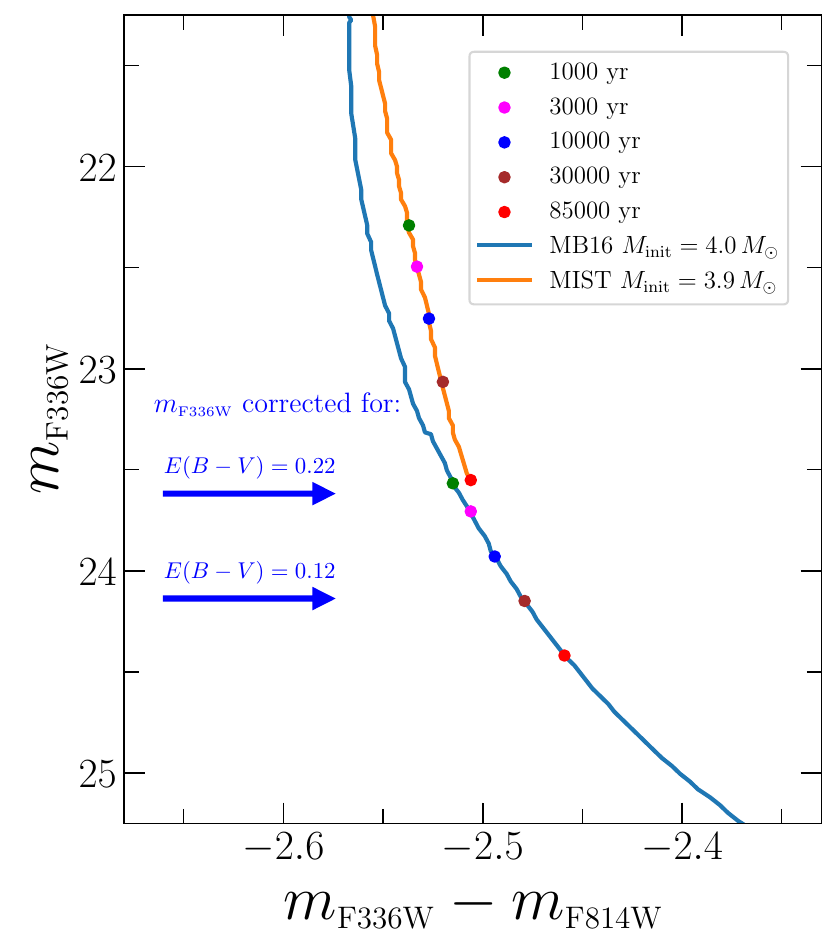}
\caption{Zoomed-in detail of the color-magnitude diagram in the vicinity of the central star. Two theoretical post-AGB evolutionary tracks are plotted, from MB16 (blue curve) and MIST (orange curve), with initial masses of 4.0 and $3.9\,M_\odot$, respectively. Post-AGB ages along the tracks are marked with filled circles, color-coded as indicated in the legend. Blue arrows mark the $m_{\rm F336W}$ levels for the central star, after corrections for a reddening of $E(B-V)=0.12$ (bottom arrow) and 0.22 (top arrow). See text for details and discussion of this figure.
\label{fig:MB2016_cmd}
}
\end{figure}

In Figure~\ref{fig:MB2016_cmd} we zoom in on the CMD of Figure~\ref{fig:cmd}, focusing on the region near the location of the PN central star. The orange line plots a MIST post-AGB evolutionary track, calculated for an initial mass of $3.9\,M_\odot$ and metallicity of $\rm[Fe/H]=-0.4$, which are parameters appropriate for NGC\,1866. The blue line plots a post-AGB track from the MB16 library. Here we chose the nearest available initial mass to that of the cluster, $4.0\,M_\odot$. The metallicity for this track is $Z=0.02$, which is unfortunately higher than that of NGC\,1866, but this discrepancy will have a relatively small effect on the tracks. 

Several post-AGB ages are marked with color-coded filled circles on the two tracks in Figure~\ref{fig:MB2016_cmd}, as indicated in the figure legend. Here, as noted above, the ages are calculated with respect to the time at which the effective temperature of the post-AGB star attains $\log(\Teff/{\rm K})=3.85$. Note that the final point on the MIST track is at an age of about 85,000~yr.

As Figure~\ref{fig:cmd} shows, the error bar for the color of the PNN is large, due to the considerable uncertainty in its F814W magnitude. However, the F336W magnitude is fairly well constrained, even considering the possible small systematic error due to the treatment of CTE (see end of Section~\ref{subsec:central_star}). We therefore use $m_{\rm F336W}$ to test theoretical predictions.

The bottom arrow on the left side of Figure~\ref{fig:MB2016_cmd} marks the F336W magnitude level of the PNN after removal of extinction corresponding to $E(B-V)=0.12$, the reddening of the host cluster. This magnitude implies a post-AGB age on the MB16 track that is reasonably consistent with the dynamical age of the PN (especially if the magnitude were brightened slightly through a CTE correction, and\slash or by assuming a small amount of dust extinction within the nebula). However, because of the slow fading rate, small uncertainties in the magnitude of the star imply large uncertainties in the implied age. The fading rate on the MIST track is considerably slower than on the MB16 track, and the PNN's magnitude level on the MIST track indicates an age that is highly inconsistent with the dynamical age.

The top arrow in the figure shows the magnitude of the PNN corrected for a larger amount of reddening, $E(B-V)=0.22$, which is implied by the analysis of the nebular spectrum described in the next section. In this case, the MB16 track would imply an age of the PNN considerably smaller than the dynamical age, but the MIST age is still much higher than $\tau_{\rm dyn}$. These results are discussed further in Section~\ref{sec:summary}.


{ Lastly, we note that the post-AGB stars described by the MB16 and MIST tracks are burning hydrogen, and they do not include complications such as late helium-shell thermal pulses. Moreover, possible interactions with a binary companion are not considered. Either of these effects could significantly alter the evolutionary timescale for the central star.}

\section{Nebular Analysis \label{sec:nebular_analysis} }

\subsection{Classification as a Type I PN}

As noted in Section~\ref{subsec:integrated_spectrum} and shown in Figure~\ref{fig:PNspectrum}, the integrated spectrum of \Ka\ exhibits very strong [\NII]  $\lambda\lambda 6548,6583$ emission lines. This suggests that the nebula belongs to the class of Type~I PNe \citep{Peimbert1978, Peimbert+Peimbert1983, Torres-Peimbert1997}. These objects have several properties in common, among which are: (1)~high He and N abundances; (2)~bipolar structure; (3)~high central-star mass, often having a high effective temperature; (4)~high nebular mass; (5)~high mass of the main-sequence progenitor star; and (6)~complex velocity fields, including high velocities. 




Regarding the morphology of \Ka, although its projected shape appears elliptical, a bipolar geometry with two faint lobes aligned to the line of sight cannot be excluded. This can be understood considering an imaginary pole-on view of, say, the bipolar Galactic PN NGC\,2346 \citep{Walsh+1991}, or of the supernova remnant N\,103B \citep{Yamaguchi+2021}. The shape of \Ka\ is remarkably similar to that of the classical Ring Nebula (NGC\,6720), which is generally considered to be a bipolar PN viewed nearly pole-on \citep[e.g.,][]{ODell2007}.

In regard to the abundances of He and N, the strength of the [\NII] lines has already been mentioned and is obvious. However, the He\,I line $\lambda 6678$ is seen at best at the detection limit, and no other He lines are present. He\,II $\lambda 4686$ would have been available in the extended mode of MUSE, which was however not used in the existing observations. Therefore, even under the assumption of a high-excitation nebula, the He abundance cannot be constrained from the present data set, and thus the classification as a Type~I PN remains somewhat ambiguous. The missing He\,II $\lambda 4686$ also means that the excitation class cannot be determined as an indicator for  central-star temperature. We have to admit that the specifics and the quality of the nebular spectrum of an object discovered serendipitously in a program with other aims, remain for now somewhat limited. 

{ However, if our progenitor mass of $3.9\, M_\odot$ is correct, it places the object well above the observed lower mass limit for hot-bottom burning, which
is responsible for the dredge-up of CNO-processed material and the formation of Type~I PNe \citep[see][and discussion therein]{Davis2019}. Thus, even in the absence of a measurement of \heii\ $\lambda$4686, we believe that the classification as a Type~I PNe is plausible. }

\subsection{Plasma Diagnostics and Extinction \label{subsec:plasma+extinction} }

Nevertheless, based on the measured emission-line fluxes of H$\alpha$, H$\beta$, [\NII] $\lambda\lambda 5755,6548$, and [\sii] $\lambda\lambda 6717,6731$, we performed plasma diagnostics using the {\tt PyNeb} code\footnote{\url{https://research.iac.es/proyecto/PyNeb}} \citep{Luridiana+2015}.  The flux uncertainties were propagated from the emission-line errors together with a 2\% flux calibration error added in quadrature \citep{Weilbacher2020}. The analysis proceeds as follows. First, extinction is determined from the Balmer decrement with a Monte Carlo approach. The line fluxes are varied randomly using a normal distribution, based on the line-flux errors, for a total of 1000 realizations. The median of the resulting distribution is taken as the extinction value, and the uncertainty taken at the 16th and 84th percentiles. Initial values of electron temperature $T_e =10000$~K and electron density $n_e = 1000\,\rm cm^{-3}$  (i.e., a standard Case~B assumption) are set at the beginning. Based on the resulting extinction value, and employing the same Monte Carlo approach, $T_e$ is determined from the [\NII] lines, and $n_e$ from the [\sii] doublet. With these new values, the extinction is determined again, and the entire sequence iterated until convergence.

Since the extinction law that should be used for these calculations is not known {\it a priori}, we experimented with two variants: the Milky Way interstellar extinction curve from \citet{Cardelli1989}, assuming an average ratio of total to selective extinction of $R_V = 3.1$;
and the LMC average curve from \citet{Gordon+2003} with $R_V = 3.41$. It is worth noting that \citet{Phillips2007} used near-infrared imaging to study two compact young PNe (NGC\,7027, BD\,$+30^\circ3639$) and two evolved PNe (Abell\,30, Abell\,78). They concluded that the dust in the compact nebulae has extinction properties similar to those of interstellar grains. However, in the evolved PNe, it appears that the extinction arises from much smaller grains, which would imply an extinction law different from that of the ISM\null.  Thus it is possible that the latter case may apply as well for the evolved PN in NGC\,1866. Table~\ref{table:extincion} presents our results based on the two choices of extinction law. Here, as usual, the logarithmic extinction at \Hb\ is denoted $c({\rm H}\beta)$.

\begin{deluxetable}{lccc}[!h]
\tablecaption{Reddening and Extinction of the PN Integrated Spectrum for Two Different Extinction Laws
\label{table:extincion} }
\tablewidth{\textwidth}
\tablehead{
\colhead{Extinction Law}
& \colhead{$E(B-V)$}
& \colhead{$c({\rm H}\beta)$}
& \colhead{$A_V$}
}
\startdata
\citet{Cardelli1989} &  $0.26 \pm 0.07$  &  $0.38 \pm 0.10$  &  $0.81 \pm 0.21$ \\
\citet{Gordon+2003}  &  $0.22 \pm 0.06$   &  $0.35 \pm 0.09$ &  $0.76 \pm 0.20$  \\
\enddata
\end{deluxetable}

The difference of only 0.05~mag in the two solutions for the visual extinction, $A_V$, is seemingly not dramatic, and is not very important for the plasma diagnostics. However, the choice of extinction curve for dereddening the central star's F336W magnitude is significant. We note that the derived values of $E(B-V)$ are larger than the 0.12 obtained from the cluster stellar photometry (see Section~\ref{subsec:cluster_members}) by 0.14 and 0.10, respectively, for the two choices of extinction law. This additional reddening can plausibly be attributed to dust within the nebula, of an amount that is not unusual for PNe. For discussions of internal extinction in PNe and its correlation with high stellar remnant masses, see, for example, \citet{Kwitter2012},   \citet{JacobyCiardullo2025}, and references therein. We discussed in the previous section whether this extra internal extinction also applies to the photometry of the PNN.

In the absence of further information, we adopted the LMC law from \citet{Gordon+2003}. On this basis, we constrain the electron temperature to $T_e = 12000 _{-900}^{+1000}$~K, and the electron density to a low value of $n_e = 110_{-70}^{+110}$~cm$^{-3}$.
Both values are in accord with a late stage of evolution, e.g., an age greater than 10,000 yrs, in the hydrodynamical PN models presented by \citet{Schonberner2005}.

\subsection{Abundances}

An estimate of the N abundance and nebular mass proves to be difficult: measuring the N abundance requires knowing the O abundance, for which in turn the ionization correction factor must be determined. The latter requires emission-line intensities for singly and doubly ionized oxygen---however, no [\oii] lines are detected in our available spectrum.

In summary, due to our limited wavelength coverage and exposure time, the plasma diagnostics are inconclusive  insofar as a clear Type~I classification is concerned---which nevertheless would be consistent with the high PNN progenitor mass inferred from the cluster age. Targeted and AO-supported deep MUSE observations of the PN in the extended mode would help resolve these issues.

\section{Summary and Future Work \label{sec:summary} }

Using the MUSE integral-field spectrograph on the ESO VLT, we have  discovered a faint PN, designated \Ka, superposed on the massive young star cluster NGC\,1866 in the LMC\null. The discovery was serendipitous, occurring in a program with a very different aim of spectroscopy of bright stars in the cluster; thus the data are less than ideal for an analysis of the PN\null.  

Nevertheless, we have reached several useful conclusions, which should pave the way for more focused observations in the future.
We show that the PN's proximity to the cluster center and the close agreement of its RV with those of cluster stars make it highly probable that the PN is a physical member of NGC\,1866. This implies that its progenitor star had an initial mass of about $3.9\,M_\odot$. Monochromatic images of the PN, derived from the MUSE data, show a ring-like morphology, { possibly} indicating a bipolar nebula viewed approximately pole-on. The integrated spectrum of the PN shows strong emission lines of [\NII]; this is consistent with it belonging to the Type~I class of PNe, arising from relatively high-mass progenitor stars.

{
\Ka\ is an old and evolved PN, as indicated by its low surface brightness, low electron density, and an apparently high dynamical age of approximately 18,000~yr.} 
We identify its probable central star in archival images obtained with the {\it Hubble Space Telescope}, allowing us to compare photometry of the star with predictions of theoretical late stellar evolution. There is reasonable agreement with the evolutionary tracks of MB16 if we assume the dynamical age, but only if the central star suffers similar interstellar extinction to that of the cluster. Our spectral analysis of the PN indicates the presence of considerable dust within the nebula; if the star were to share this higher amount of extinction, then its luminosity would be incompatible with MB16. This may indicate that we are viewing the central star through a low-extinction window from a pole-on viewpoint; { however, this remains only a speculation in the absence of actual kinematic data for the nebula.}

There are now a half dozen PNe that appear to be members of OCs---although some of them have not yet had measurements of the proper motions of their central stars that would confirm their membership. We list these six PNe in Table~\ref{table:PNe_in_clusters}. Column~3 lists the progenitor masses for these PNe, covering a range of
2.1 to $5.6\,M_\odot$. As was noted recently by \citet{BelliniPHR2025}, these high masses are strikingly discordant with statistical arguments by \citet{Badenes2015} that a large majority of field PNe in the LMC had initial masses of 1.0 to $1.2\,M_\odot$, with only a small fraction arising from more massive stars. A possible explanation for the higher masses that dominate the progenitors of PNe found in clusters may be that most OCs dissipate into the field on a timescale shorter than the evolutionary lifetimes of their relatively low-mass stars. Thus the PNe that are discovered in OCs will preferentially arise from higher-mass progenitors lying in the minority population of younger clusters that are still intact. 

\begin{deluxetable*}{lccc}
\tablecaption{Planetary Nebulae Belonging to Open Star Clusters
\label{table:PNe_in_clusters} }
\tablewidth{\textwidth}
\tablehead{
\colhead{PN}
& \colhead{Cluster}
& \colhead{Progenitor}
& \colhead{References\tablenotemark{a}}\\
\colhead{}
& \colhead{}
& \colhead{Mass [$M_\odot$]}
& \colhead{}
}
\startdata
BMP\,J1613$-$6555       & NGC\,6067       & 5.6 & (1) \\
Ka LMC 1                & NGC\,1866 (LMC) & 3.9 & (2) \\
M31 B477-1              & B477-D075 (M31) & 3.4 & (3) \\
IPHASX J055226.2+323724 & M37             & 2.8 & (4) \\
Hb 2                    & NGC\,2818       & 2.3 & (5) \\
PHR\,J1315$-$6555       & AL 1            & 2.1 & (6) \\
\enddata
\tablenotetext{a}{References for progenitor masses: (1) \citet{FragkouNGC6067_2022}; (2)  This paper; (3) \citet{Davis2019}; (4) \citet{Griggio2022, FragkouM37_2022, WernerM372023}; (5) \citet{Fragkou2025}; however cluster membership is dubious (see our Section~\ref{sec:intro}); (6)~\citet{FragkouPHR2019, BelliniPHR2025}.  }
\end{deluxetable*}

It is no surprise that the fortuitous nature of the discovery of \Ka, and the consequent limitations of the available data, impose shortcomings on our initial analysis: (1)~The PN is extremely faint: it lies 8~magnitudes below the bright end of the LMC PNLF\null. The MUSE exposure time, which was sufficient for the spectroscopic study of the cluster stars, was too short to reach important diagnostic emission lines that would be needed for an abundance determination and an estimate of the nebular mass. (2)~The nominal MUSE wavelength range excludes important diagnostic lines in the blue, preventing accurate determinations of the excitation class of the nebula, and, again, chemical abundances. (3)~The MUSE spectral resolution is too coarse to measure the nebular expansion velocity. (4)~The angular resolution from ground-based observations with MUSE is an order of magnitude inferior to that which can be obtained with \HST, thus rendering the removal of stellar residuals from nebular emission-line maps difficult. (5)~It remains unclear whether the determination of extinction obtained from the Balmer decrement in the bright nebular rim, i.e., away from the center, is applicable for the central star reddening. (6)~While the \HST/WFC3 photometry in F336W comes with a reasonable error, the F814W magnitude is extremely uncertain, and this is a severe handicap when trying to constrain the reddening of the central star. (7) There are no \HST/WFC3 images covering a long enough epoch to estimate a proper motion of the PNN, which would provide further evidence, along with the radial velocity, of cluster membership.

{
These limitations could be addressed with future observing campaigns. MUSE
NFM (narrow-field mode) and the upcoming MAVIS would improve the spatial
resolution to study the geometry better. Deep spectroscopy, preferably at
medium to high resolution and at shorter wavelengths, would offer a better understanding of the kinematics, along with access to the important diagnostic lines, including  [\oii] $\lambda$3727-29 and \heii\ $\lambda$4686.
These lines are crucial for determining the elemental abundances of helium,
oxygen, and nitrogen, and could provide a strong confirmation of the classification as a Type~I PN\null. Such spectroscopy would also enable a measurement of the expansion velocity of the PN, providing additional constraints on its geometry and dynamical age. There is also a crucial need for deeper multi-color optical and UV imaging with
\HST, which would provide much more precise photometry for the central star,
and tighter constraints on its extinction.}

Lastly, our comparison with theoretical evolutionary tracks suffered from the limited number of published studies using state-of-the-art physics, such as the MB16 library. The discrepancies between the post-AGB tracks from two modern and widely used tools, as shown in Figure~\ref{fig:MB2016_cmd}, emphasizes the continuing need for theoretical studies of this still poorly understood area of stellar astrophysics.


\acknowledgments

Based on observations collected at the European Southern Observatory under ESO programme 0104.D-0257(B). S.K. gratefully acknowledges funding from UKRI through a Future Leaders Fellowship (grant MR/Y034147/1).

Based in part on observations with the NASA/ESA {\it Hubble Space Telescope\/} obtained from the Data Archive at the Space Telescope Science Institute (STScI), which is operated by the Association of Universities for Research in Astronomy, Incorporated, under NASA contract NAS5-26555. 
Support for Program number GO-16748 was provided through grants from STScI under NASA contract NAS5-26555.

This work has made use of data from the European Space Agency (ESA) mission{\it Gaia\/} (\url{https://www.cosmos.esa.int/gaia}), processed by the {\it Gaia\/} Data Processing and Analysis Consortium (DPAC, \url{https://www.cosmos.esa.int/web/gaia/dpac/consortium}). Funding for the DPAC has been provided by national institutions, in particular the institutions participating in the {\it Gaia\/} Multilateral Agreement.

F.N. acknowledges funding by DLR grant 50 OR 2216.  

M.M.R. acknowledges support from BMFTR under grant 03WSP1745.

We thank Marcelo Miller Bertolami and Aaron Dotter for useful comments.






\bibliography{PNNisurvey_refs}

\begin{thebibliography}{}
\expandafter\ifx\csname natexlab\endcsname\relax\def\natexlab#1{#1}\fi
\providecommand{\url}[1]{\href{#1}{#1}}

\bibitem[{{Anderson}(2021)}]{Anderson2021}
{Anderson}, J. 2021, {Table-Based CTE Corrections for flt-Format WFC3/UVIS}, Instrument Science Report WFC3 2021-13, ,

\bibitem[{{Bacon} {et~al.}(2014){Bacon}, {Vernet}, {Borisova}, {Bouch{\'e}}, {Brinchmann}, {Carollo}, {Carton}, {Caruana}, {Cerda}, {Contini}, {Franx}, {Girard}, {Guerou}, {Haddad}, {Hau}, {Herenz}, {Herrera}, {Husemann}, {Husser}, {Jarno}, {Kamann}, {Krajnovic}, {Lilly}, {Mainieri}, {Martinsson}, {Palsa}, {Patricio}, {P{\'e}contal}, {Pello}, {Piqueras}, {Richard}, {Sandin}, {Schroetter}, {Selman}, {Shirazi}, {Smette}, {Soto}, {Streicher}, {Urrutia}, {Weilbacher}, {Wisotzki}, \& {Zins}}]{Bacon2014}
{Bacon}, R., {Vernet}, J., {Borisova}, E., {et~al.} 2014, The Messenger, 157, 13

\bibitem[{{Badenes} {et~al.}(2015){Badenes}, {Maoz}, \& {Ciardullo}}]{Badenes2015}
{Badenes}, C., {Maoz}, D., \& {Ciardullo}, R. 2015, \apjl, 804, L25

\bibitem[{{Bastian} \& {de Mink}(2009)}]{Bastian2009}
{Bastian}, N., \& {de Mink}, S.~E. 2009, \mnras, 398, L11

\bibitem[{{Bellini} {et~al.}(2017){Bellini}, {Anderson}, {Bedin}, {King}, {van der Marel}, {Piotto}, \& {Cool}}]{Bellini2017}
{Bellini}, A., {Anderson}, J., {Bedin}, L.~R., {et~al.} 2017, \apj, 842, 6

\bibitem[{{Bellini} {et~al.}(2025){Bellini}, {Bond}, \& {Sahu}}]{BelliniPHR2025}
{Bellini}, A., {Bond}, H.~E., \& {Sahu}, K.~C. 2025, \aj, 169, 199

\bibitem[{{Bohlin} {et~al.}(2020){Bohlin}, {Hubeny}, \& {Rauch}}]{Bohlin2020}
{Bohlin}, R.~C., {Hubeny}, I., \& {Rauch}, T. 2020, \aj, 160, 21

\bibitem[{{Boji{\v{c}}i{\'c}} {et~al.}(2017){Boji{\v{c}}i{\'c}}, {Parker}, \& {Frew}}]{Bojicic2017}
{Boji{\v{c}}i{\'c}}, I.~S., {Parker}, Q.~A., \& {Frew}, D.~J. 2017, in IAU Symposium, Vol. 323, Planetary Nebulae: Multi-Wavelength Probes of Stellar and Galactic Evolution, ed. X.~{Liu}, L.~{Stanghellini}, \& A.~{Karakas}, 327--328

\bibitem[{{Bond}(2015)}]{Bond2015}
{Bond}, H.~E. 2015, \aj, 149, 132

\bibitem[{{Bond} {et~al.}(2020){Bond}, {Bellini}, \& {Sahu}}]{Bond2020}
{Bond}, H.~E., {Bellini}, A., \& {Sahu}, K.~C. 2020, \aj, 159, 276

\bibitem[{{Bond} {et~al.}(2024){Bond}, {Bellini}, \& {Sahu}}]{Bond_JaFu12024}
---. 2024, \aj, 168, 160

\bibitem[{{Calamida} {et~al.}(2022){Calamida}, {Bajaj}, {Mack}, {Marinelli}, {Medina}, {Pidgeon}, {Kozhurina-Platais}, {Shanahan}, \& {Som}}]{Calamida2022}
{Calamida}, A., {Bajaj}, V., {Mack}, J., {et~al.} 2022, \aj, 164, 32

\bibitem[{{Cardelli} {et~al.}(1989){Cardelli}, {Clayton}, \& {Mathis}}]{Cardelli1989}
{Cardelli}, J.~A., {Clayton}, G.~C., \& {Mathis}, J.~S. 1989, \apj, 345, 245

\bibitem[{{Choi} {et~al.}(2016){Choi}, {Dotter}, {Conroy}, {Cantiello}, {Paxton}, \& {Johnson}}]{Choi2016}
{Choi}, J., {Dotter}, A., {Conroy}, C., {et~al.} 2016, \apj, 823, 102

\bibitem[{{Costa} {et~al.}(2019){Costa}, {Girardi}, {Bressan}, {Chen}, {Goudfrooij}, {Marigo}, {Rodrigues}, \& {Lanza}}]{Costa2019}
{Costa}, G., {Girardi}, L., {Bressan}, A., {et~al.} 2019, \aap, 631, A128

\bibitem[{{Davis} {et~al.}(2019){Davis}, {Bond}, {Ciardullo}, \& {Jacoby}}]{Davis2019}
{Davis}, B.~D., {Bond}, H.~E., {Ciardullo}, R., \& {Jacoby}, G.~H. 2019, \apj, 884, 115

\bibitem[{{Dotter}(2016)}]{Dotter2016}
{Dotter}, A. 2016, \apjs, 222, 8

\bibitem[{{Ettorre} {et~al.}(2025){Ettorre}, {Mazzi}, {Girardi}, {Marigo}, {Pastorelli}, {Goudfrooij}, {Williams}, {Bellini}, {Bressan}, {Chen}, {Correnti}, {Costa}, {Dalcanton}, {Facchini}, {Fouesneau}, {Nguyen}, \& {Volpato}}]{Ettorre2025}
{Ettorre}, G., {Mazzi}, A., {Girardi}, L., {et~al.} 2025, \mnras, 539, 2537

\bibitem[{{Foreman-Mackey} {et~al.}(2013){Foreman-Mackey}, {Hogg}, {Lang}, \& {Goodman}}]{2013PASP..125..306F}
{Foreman-Mackey}, D., {Hogg}, D.~W., {Lang}, D., \& {Goodman}, J. 2013, \pasp, 125, 306

\bibitem[{{Fragkou} {et~al.}(2019{\natexlab{a}}){Fragkou}, {Parker}, {Zijlstra}, {Shaw}, \& {Lykou}}]{FragkouPHR2019}
{Fragkou}, V., {Parker}, Q.~A., {Zijlstra}, A., {Shaw}, R., \& {Lykou}, F. 2019{\natexlab{a}}, \mnras, 484, 3078

\bibitem[{{Fragkou} {et~al.}(2019{\natexlab{b}}){Fragkou}, {Parker}, {Zijlstra}, {Crause}, \& {Barker}}]{FragkouBMP2019}
{Fragkou}, V., {Parker}, Q.~A., {Zijlstra}, A.~A., {Crause}, L., \& {Barker}, H. 2019{\natexlab{b}}, Nature Astronomy, 3, 851

\bibitem[{{Fragkou} {et~al.}(2022{\natexlab{a}}){Fragkou}, {Parker}, {Zijlstra}, {Crause}, {Sabin}, \& {V{\'a}zquez}}]{FragkouNGC6067_2022}
{Fragkou}, V., {Parker}, Q.~A., {Zijlstra}, A.~A., {et~al.} 2022{\natexlab{a}}, Galaxies, 10, 44

\bibitem[{{Fragkou} {et~al.}(2022{\natexlab{b}}){Fragkou}, {Parker}, {Zijlstra}, {V{\'a}zquez}, {Sabin}, \& {Rechy-Garcia}}]{FragkouM37_2022}
---. 2022{\natexlab{b}}, \apjl, 935, L35

\bibitem[{{Fragkou} {et~al.}(2025){Fragkou}, {V{\'a}zquez}, {Parker}, {Gon{\c{c}}alves}, \& {Lomel{\'\i}-N{\'u}{\~n}ez}}]{Fragkou2025}
{Fragkou}, V., {V{\'a}zquez}, R., {Parker}, Q.~A., {Gon{\c{c}}alves}, D.~R., \& {Lomel{\'\i}-N{\'u}{\~n}ez}, L. 2025, \aap, 696, A146

\bibitem[{{Gaia Collaboration} {et~al.}(2023){Gaia Collaboration}, {Vallenari}, {Brown}, {Prusti}, {de Bruijne}, {Arenou}, {Babusiaux}, {Biermann}, {Creevey}, {Ducourant}, {Evans}, {Eyer}, {Guerra}, {Hutton}, {Jordi}, {Klioner}, {Lammers}, {Lindegren}, {Luri}, {Mignard}, {Panem}, {Pourbaix}, {Randich}, {Sartoretti}, {Soubiran}, {Tanga}, {Walton}, {Bailer-Jones}, {Bastian}, {Drimmel}, {Jansen}, {Katz}, {Lattanzi}, {van Leeuwen}, {Bakker}, {Cacciari}, {Casta{\~n}eda}, {De Angeli}, {Fabricius}, {Fouesneau}, {Fr{\'e}mat}, {Galluccio}, {Guerrier}, {Heiter}, {Masana}, {Messineo}, {Mowlavi}, {Nicolas}, {Nienartowicz}, {Pailler}, {Panuzzo}, {Riclet}, {Roux}, {Seabroke}, {Sordo}, {Th{\'e}venin}, {Gracia-Abril}, {Portell}, {Teyssier}, {Altmann}, {Andrae}, {Audard}, {Bellas-Velidis}, {Benson}, {Berthier}, {Blomme}, {Burgess}, {Busonero}, {Busso}, {C{\'a}novas}, {Carry}, {Cellino}, {Cheek}, {Clementini}, {Damerdji}, {Davidson}, {de Teodoro}, {Nu{\~n}ez Campos}, {Delchambre}, {Dell'Oro}, {Esquej},
  {Fern{\'a}ndez-Hern{\'a}ndez}, {Fraile}, {Garabato}, {Garc{\'\i}a-Lario}, {Gosset}, {Haigron}, {Halbwachs}, {Hambly}, {Harrison}, {Hern{\'a}ndez}, {Hestroffer}, {Hodgkin}, {Holl}, {Jan{\ss}en}, {Jevardat de Fombelle}, {Jordan}, {Krone-Martins}, {Lanzafame}, {L{\"o}ffler}, {Marchal}, {Marrese}, {Moitinho}, {Muinonen}, {Osborne}, {Pancino}, {Pauwels}, {Recio-Blanco}, {Reyl{\'e}}, {Riello}, {Rimoldini}, {Roegiers}, {Rybizki}, {Sarro}, {Siopis}, {Smith}, {Sozzetti}, {Utrilla}, {van Leeuwen}, {Abbas}, {{\'A}brah{\'a}m}, {Abreu Aramburu}, {Aerts}, {Aguado}, {Ajaj}, {Aldea-Montero}, {Altavilla}, {{\'A}lvarez}, {Alves}, {Anders}, {Anderson}, {Anglada Varela}, {Antoja}, {Baines}, {Baker}, {Balaguer-N{\'u}{\~n}ez}, {Balbinot}, {Balog}, {Barache}, {Barbato}, {Barros}, {Barstow}, {Bartolom{\'e}}, {Bassilana}, {Bauchet}, {Becciani}, {Bellazzini}, {Berihuete}, {Bernet}, {Bertone}, {Bianchi}, {Binnenfeld}, {Blanco-Cuaresma}, {Blazere}, {Boch}, {Bombrun}, {Bossini}, {Bouquillon}, {Bragaglia}, {Bramante}, {Breedt},
  {Bressan}, {Brouillet}, {Brugaletta}, {Bucciarelli}, {Burlacu}, {Butkevich}, {Buzzi}, {Caffau}, {Cancelliere}, {Cantat-Gaudin}, {Carballo}, {Carlucci}, {Carnerero}, {Carrasco}, {Casamiquela}, {Castellani}, {Castro-Ginard}, {Chaoul}, {Charlot}, {Chemin}, {Chiaramida}, {Chiavassa}, {Chornay}, {Comoretto}, {Contursi}, {Cooper}, {Cornez}, {Cowell}, {Crifo}, {Cropper}, {Crosta}, {Crowley}, {Dafonte}, {Dapergolas}, {David}, {David}, {de Laverny}, {De Luise}, \& {De March}}]{2023A&A...674A...1G}
{Gaia Collaboration}, {Vallenari}, A., {Brown}, A.~G.~A., {et~al.} 2023, \aap, 674, A1

\bibitem[{{Gordon} {et~al.}(2003){Gordon}, {Clayton}, {Misselt}, {Landolt}, \& {Wolff}}]{Gordon+2003}
{Gordon}, K.~D., {Clayton}, G.~C., {Misselt}, K.~A., {Landolt}, A.~U., \& {Wolff}, M.~J. 2003, \apj, 594, 279

\bibitem[{{G{\"o}ttgens} {et~al.}(2019){G{\"o}ttgens}, {Weilbacher}, {Roth}, {Dreizler}, {Giesers}, {Husser}, {Kamann}, {Brinchmann}, {Kollatschny}, {Monreal-Ibero}, {Schmidt}, {Wendt}, {Wisotzki}, \& {Bacon}}]{Goettgens2019}
{G{\"o}ttgens}, F., {Weilbacher}, P.~M., {Roth}, M.~M., {et~al.} 2019, \aap, 626, A69

\bibitem[{{Griggio} {et~al.}(2022){Griggio}, {Bedin}, {Raddi}, {Reindl}, {Tomasella}, {Scalco}, {Salaris}, {Cassisi}, {Ochner}, {Ciroi}, {Rosati}, {Nardiello}, {Anderson}, {Libralato}, {Bellini}, {Vallenari}, {Spina}, \& {Pedani}}]{Griggio2022}
{Griggio}, M., {Bedin}, L.~R., {Raddi}, R., {et~al.} 2022, \mnras, 515, 1841

\bibitem[{{Henry} {et~al.}(2018){Henry}, {Stephenson}, {Miller Bertolami}, {Kwitter}, \& {Balick}}]{Henry2018}
{Henry}, R.~B.~C., {Stephenson}, B.~G., {Miller Bertolami}, M.~M., {Kwitter}, K.~B., \& {Balick}, B. 2018, \mnras, 473, 241

\bibitem[{{Hubble}(1921)}]{Hubble1921}
{Hubble}, E. 1921, \pasp, 33, 174

\bibitem[{{Jacoby}(1989)}]{Jacoby1989}
{Jacoby}, G.~H. 1989, \apj, 339, 39

\bibitem[{{Jacoby} \& {Ciardullo}(2025)}]{JacobyCiardullo2025}
{Jacoby}, G.~H., \& {Ciardullo}, R. 2025, \apj, 983, 129

\bibitem[{{Jacoby} {et~al.}(2024){Jacoby}, {Ciardullo}, {Roth}, {Arnaboldi}, \& {Weilbacher}}]{Jacoby+2024}
{Jacoby}, G.~H., {Ciardullo}, R., {Roth}, M.~M., {Arnaboldi}, M., \& {Weilbacher}, P.~M. 2024, \apjs, 271, 40

\bibitem[{{Kamann} {et~al.}(2013){Kamann}, {Wisotzki}, \& {Roth}}]{Kamann+2013}
{Kamann}, S., {Wisotzki}, L., \& {Roth}, M.~M. 2013, \aap, 549, A71

\bibitem[{{Kamann} {et~al.}(2020){Kamann}, {Bastian}, {Gossage}, {Baade}, {Cabrera-Ziri}, {Da Costa}, {de Mink}, {Georgy}, {Giesers}, {G{\"o}ttgens}, {Hilker}, {Husser}, {Lardo}, {Larsen}, {Mackey}, {Martocchia}, {Mucciarelli}, {Platais}, {Roth}, {Salaris}, {Usher}, \& {Yong}}]{Kamann2020}
{Kamann}, S., {Bastian}, N., {Gossage}, S., {et~al.} 2020, \mnras, 492, 2177

\bibitem[{{Kamann} {et~al.}(2023){Kamann}, {Saracino}, {Bastian}, {Gossage}, {Usher}, {Baade}, {Cabrera-Ziri}, {de Mink}, {Ekstrom}, {Georgy}, {Hilker}, {Larsen}, {Mackey}, {Niederhofer}, {Platais}, \& {Yong}}]{Kamann2023}
{Kamann}, S., {Saracino}, S., {Bastian}, N., {et~al.} 2023, \mnras, 518, 1505

\bibitem[{{Kamann} {et~al.}(2025){Kamann}, {Bastian}, {Niederhofer}, {Bellini}, {Cabrera-Ziri}, {Dreizler}, {G{\"o}ttgens}, {Kozhurina-Platais}, {Libralato}, {Martens}, \& {Saracino}}]{Kamann2025}
{Kamann}, S., {Bastian}, N., {Niederhofer}, F., {et~al.} 2025, \mnras, 542, 2768, (K25)

\bibitem[{{Kamath} {et~al.}(2023){Kamath}, {Dell'Agli}, {Ventura}, {Van Winckel}, {Tosi}, \& {Karakas}}]{Kamath2023}
{Kamath}, D., {Dell'Agli}, F., {Ventura}, P., {et~al.} 2023, \mnras, 519, 2169

\bibitem[{{Karakas} {et~al.}(2018){Karakas}, {Lugaro}, {Carlos}, {Cseh}, {Kamath}, \& {Garc{\'\i}a-Hern{\'a}ndez}}]{Karakas2018}
{Karakas}, A.~I., {Lugaro}, M., {Carlos}, M., {et~al.} 2018, \mnras, 477, 421

\bibitem[{{Kuhn}(2024)}]{Kuhn2024}
{Kuhn}, B. 2024, {WFC3/UVIS External CTE Monitoring 2009-2024}, Instrument Science Report WFC3 2024-04, 16 pages, ,

\bibitem[{{Kwitter} \& {Henry}(2022)}]{Kwitter2022}
{Kwitter}, K.~B., \& {Henry}, R.~B.~C. 2022, \pasp, 134, 022001

\bibitem[{{Kwitter} {et~al.}(2012){Kwitter}, {Lehman}, {Balick}, \& {Henry}}]{Kwitter2012}
{Kwitter}, K.~B., {Lehman}, E. M.~M., {Balick}, B., \& {Henry}, R.~B.~C. 2012, \apj, 753, 12

\bibitem[{{Kwok}(2022)}]{Kwok2022}
{Kwok}, S. 2022, Frontiers in Astronomy and Space Sciences, 9, 893061

\bibitem[{{Larsen} \& {Richtler}(2006)}]{Larsen2006}
{Larsen}, S.~S., \& {Richtler}, T. 2006, \aap, 459, 103

\bibitem[{{Luridiana} {et~al.}(2015){Luridiana}, {Morisset}, \& {Shaw}}]{Luridiana+2015}
{Luridiana}, V., {Morisset}, C., \& {Shaw}, R.~A. 2015, \aap, 573, A42

\bibitem[{{Martens} {et~al.}(2023){Martens}, {Kamann}, {Dreizler}, {G{\"o}ttgens}, {Husser}, {Latour}, {Balakina}, {Krajnovi{\'c}}, {Pechetti}, \& {Weilbacher}}]{2023A&A...671A.106M}
{Martens}, S., {Kamann}, S., {Dreizler}, S., {et~al.} 2023, \aap, 671, A106

\bibitem[{{McLaughlin} \& {van der Marel}(2005)}]{McLauoghlin2005}
{McLaughlin}, D.~E., \& {van der Marel}, R.~P. 2005, \apjs, 161, 304

\bibitem[{{Miller Bertolami}(2016)}]{MillerBertolami2016}
{Miller Bertolami}, M.~M. 2016, \aap, 588, A25, (MB16)

\bibitem[{{Milone} {et~al.}(2017){Milone}, {Marino}, {D'Antona}, {Bedin}, {Piotto}, {Jerjen}, {Anderson}, {Dotter}, {di Criscienzo}, \& {Lagioia}}]{Milone2017}
{Milone}, A.~P., {Marino}, A.~F., {D'Antona}, F., {et~al.} 2017, \mnras, 465, 4363

\bibitem[{{Milone} {et~al.}(2018){Milone}, {Marino}, {Di Criscienzo}, {D'Antona}, {Bedin}, {Da Costa}, {Piotto}, {Tailo}, {Dotter}, {Angeloni}, {Anderson}, {Jerjen}, {Li}, {Dupree}, {Granata}, {Lagioia}, {Mackey}, {Nardiello}, \& {Vesperini}}]{Milone2018}
{Milone}, A.~P., {Marino}, A.~F., {Di Criscienzo}, M., {et~al.} 2018, \mnras, 477, 2640

\bibitem[{{Milone} {et~al.}(2023){Milone}, {Cordoni}, {Marino}, {D'Antona}, {Bellini}, {Di Criscienzo}, {Dondoglio}, {Lagioia}, {Langer}, {Legnardi}, {Libralato}, {Baumgardt}, {Bettinelli}, {Cavecchi}, {de Grijs}, {Deng}, {Hastings}, {Li}, {Mohandasan}, {Renzini}, {Vesperini}, {Wang}, {Ziliotto}, {Carlos}, {Costa}, {Dell'Agli}, {Di Stefano}, {Jang}, {Martorano}, {Simioni}, {Tailo}, \& {Ventura}}]{Milone2023}
{Milone}, A.~P., {Cordoni}, G., {Marino}, A.~F., {et~al.} 2023, \aap, 672, A161

\bibitem[{{Moni Bidin} {et~al.}(2014){Moni Bidin}, {Majaess}, {Bonatto}, {Mauro}, {Turner}, {Geisler}, {Chen{\'e}}, {Gormaz-Matamala}, {Borissova}, {Kurtev}, {Minniti}, {Carraro}, \& {Gieren}}]{MoniBidin2014}
{Moni Bidin}, C., {Majaess}, D., {Bonatto}, C., {et~al.} 2014, \aap, 561, A119

\bibitem[{{Niederhofer} {et~al.}(2024){Niederhofer}, {Bellini}, {Kozhurina-Platais}, {Libralato}, {H{\"a}berle}, {Kacharov}, {Kamann}, {Bastian}, {Cabrera-Ziri}, {Cioni}, {Dresbach}, {Martocchia}, {Massari}, \& {Saracino}}]{Niederhofer2024}
{Niederhofer}, F., {Bellini}, A., {Kozhurina-Platais}, V., {et~al.} 2024, \aap, 689, A162, (N24)

\bibitem[{{O'Dell} {et~al.}(2007){O'Dell}, {Sabbadin}, \& {Henney}}]{ODell2007}
{O'Dell}, C.~R., {Sabbadin}, F., \& {Henney}, W.~J. 2007, \aj, 134, 1679

\bibitem[{{Parker}(2022)}]{Parker2022}
{Parker}, Q.~A. 2022, Frontiers in Astronomy and Space Sciences, 9, 895287

\bibitem[{{Parker} {et~al.}(2016){Parker}, {Boji{\v{c}}i{\'c}}, \& {Frew}}]{Parker2016}
{Parker}, Q.~A., {Boji{\v{c}}i{\'c}}, I.~S., \& {Frew}, D.~J. 2016, in Journal of Physics Conference Series, Vol. 728, Journal of Physics Conference Series, 032008

\bibitem[{{Paxton} {et~al.}(2011){Paxton}, {Bildsten}, {Dotter}, {Herwig}, {Lesaffre}, \& {Timmes}}]{Paxton2011}
{Paxton}, B., {Bildsten}, L., {Dotter}, A., {et~al.} 2011, \apjs, 192, 3

\bibitem[{{Peimbert}(1978)}]{Peimbert1978}
{Peimbert}, M. 1978, in IAU Symposium, Vol.~76, Planetary Nebulae, ed. Y.~{Terzian}, 215--224

\bibitem[{{Peimbert} \& {Torres-Peimbert}(1983)}]{Peimbert+Peimbert1983}
{Peimbert}, M., \& {Torres-Peimbert}, S. 1983, in IAU Symposium, Vol. 103, Planetary Nebulae, ed. L.~H. {Aller}, 233--242

\bibitem[{{Phillips} \& {Ramos-Larios}(2007)}]{Phillips2007}
{Phillips}, J.~P., \& {Ramos-Larios}, G. 2007, \aj, 133, 347

\bibitem[{{Plummer}(1911)}]{1911MNRAS..71..460P}
{Plummer}, H.~C. 1911, \mnras, 71, 460

\bibitem[{{Reid} \& {Parker}(2010)}]{Reid+2010}
{Reid}, W.~A., \& {Parker}, Q.~A. 2010, \mnras, 405, 1349

\bibitem[{{Reindl} {et~al.}(2024){Reindl}, {Bond}, {Werner}, \& {Zeimann}}]{Reindl+2024}
{Reindl}, N., {Bond}, H.~E., {Werner}, K., \& {Zeimann}, G.~R. 2024, \aap, 690, A366

\bibitem[{{Roth} {et~al.}(2004){Roth}, {Becker}, {Kelz}, \& {Schmoll}}]{Roth+2004}
{Roth}, M.~M., {Becker}, T., {Kelz}, A., \& {Schmoll}, J. 2004, \apj, 603, 531

\bibitem[{{Roth} {et~al.}(2021){Roth}, {Jacoby}, {Ciardullo}, {Davis}, {Chase}, \& {Weilbacher}}]{Roth+2021}
{Roth}, M.~M., {Jacoby}, G.~H., {Ciardullo}, R., {et~al.} 2021, \apj, 916, 21

\bibitem[{{Roth} {et~al.}(2018){Roth}, {Sandin}, {Kamann}, {Husser}, {Weilbacher}, {Monreal-Ibero}, {Bacon}, {den Brok}, {Dreizler}, {Kelz}, {Marino}, \& {Steinmetz}}]{Roth+2018}
{Roth}, M.~M., {Sandin}, C., {Kamann}, S., {et~al.} 2018, \aap, 618, A3

\bibitem[{{Sch{\"o}nberner} {et~al.}(2005{\natexlab{a}}){Sch{\"o}nberner}, {Jacob}, \& {Steffen}}]{Schonberner2005b}
{Sch{\"o}nberner}, D., {Jacob}, R., \& {Steffen}, M. 2005{\natexlab{a}}, \aap, 441, 573

\bibitem[{{Sch{\"o}nberner} {et~al.}(2005{\natexlab{b}}){Sch{\"o}nberner}, {Jacob}, {Steffen}, {Perinotto}, {Corradi}, \& {Acker}}]{Schonberner2005}
{Sch{\"o}nberner}, D., {Jacob}, R., {Steffen}, M., {et~al.} 2005{\natexlab{b}}, \aap, 431, 963

\bibitem[{{Stetson}(1987)}]{Stetson1987}
{Stetson}, P.~B. 1987, \pasp, 99, 191

\bibitem[{{Torres-Peimbert} \& {Peimbert}(1997)}]{Torres-Peimbert1997}
{Torres-Peimbert}, S., \& {Peimbert}, M. 1997, in IAU Symposium, Vol. 180, Planetary Nebulae, ed. H.~J. {Habing} \& H.~J.~G.~L.~M. {Lamers}, 175

\bibitem[{{Walsh} {et~al.}(1991){Walsh}, {Meaburn}, \& {Whitehead}}]{Walsh+1991}
{Walsh}, J.~R., {Meaburn}, J., \& {Whitehead}, M.~J. 1991, \aap, 248, 613

\bibitem[{{Wang} {et~al.}(2023){Wang}, {Hastings}, {Schootemeijer}, {Langer}, {de Mink}, {Bodensteiner}, {Milone}, {Justham}, \& {Marchant}}]{Wang2023}
{Wang}, C., {Hastings}, B., {Schootemeijer}, A., {et~al.} 2023, \aap, 670, A43

\bibitem[{{Weilbacher} {et~al.}(2020){Weilbacher}, {Palsa}, {Streicher}, {Bacon}, {Urrutia}, {Wisotzki}, {Conseil}, {Husemann}, {Jarno}, {Kelz}, {P{\'e}contal-Rousset}, {Richard}, {Roth}, {Selman}, \& {Vernet}}]{Weilbacher2020}
{Weilbacher}, P.~M., {Palsa}, R., {Streicher}, O., {et~al.} 2020, \aap, 641, A28

\bibitem[{{Werner} {et~al.}(2023){Werner}, {Reindl}, {Raddi}, {Griggio}, {Bedin}, {Camisassa}, {Rebassa-Mansergas}, {Torres}, \& {Goodhew}}]{WernerM372023}
{Werner}, K., {Reindl}, N., {Raddi}, R., {et~al.} 2023, \aap, 678, A89

\bibitem[{{Yamaguchi} {et~al.}(2021){Yamaguchi}, {Acero}, {Li}, \& {Chu}}]{Yamaguchi+2021}
{Yamaguchi}, H., {Acero}, F., {Li}, C.-J., \& {Chu}, Y.-H. 2021, \apjl, 910, L24

\end{thebibliography}

\end{document}